\documentclass[fleqn,usenatbib]{mnras}

\usepackage{newtxtext,newtxmath}

\usepackage[T1]{fontenc}
\usepackage{ae,aecompl}
\usepackage[]{algorithmic}
\usepackage[linesnumbered,lined,boxed,commentsnumbered]{algorithm2e}

\usepackage{graphicx}	
\usepackage{amsmath}	
\usepackage{amssymb}	
\usepackage{multicol}        
\usepackage{bm}		
\usepackage{pdflscape}	
\usepackage{threeparttable}

\newcommand{\ms}{m\,s$^{-1}$}
\newcommand{\kms}{km\,s$^{-1}$}

\newcommand{\fortran}{\texttt{FORTRAN}}
\newcommand{\python}{\texttt{PYTHON}}

\title[\texttt{pyaneti}]{\texttt{pyaneti}: a fast and powerful software suite for multi-planet radial velocity and transit fitting}

\author[Barrag\'an et al.]{
O. Barrag\'an,$^{1}$\thanks{E-mail: oscar.barraganvil@edu.unito.it}
D. Gandolfi,$^{1}$
and G. Antoniciello$^{1, 2}$
\\
$^{1}$Dipartimento di Fisica, Universit\`{a} di Torino, via P. Giuria 1, 10125 Torino, Italy\\
$^{2}$Dipartimento di Fisica e Astronomia ``Galileo Galilei'', Universit\`a di Padova, Vicolo dell'Osservatorio 3, 35122 Padova, Italy \\
}

\date{Accepted XXX. Received YYY; in original form ZZZ}

\pubyear{2018}

\begin{document}
\label{firstpage}
\pagerange{\pageref{firstpage}--\pageref{lastpage}}
\maketitle

\begin{abstract}
Transiting exoplanet parameter estimation from time-series photometry and Doppler spectroscopy is fundamental to study planets' internal structures and compositions. Here we present the code \texttt{pyaneti}, a powerful and user-friendly software suite to perform multi-planet radial velocity and transit data fitting. The code uses a Bayesian approach combined with an MCMC sampling to estimate the parameters of planetary systems. We combine the numerical efficiency of \fortran, the versatility of \python, and the parallelization of \texttt{OpenMP} to make \texttt{pyaneti} a fast and easy to use code. The package is freely available at \url{https://github.com/oscaribv/pyaneti}.
\newline

%
\end{abstract}




\section{Introduction}
\label{sec:intro}

A new branch of astrophysics was born with the discovery of the first planet orbiting a star other than the Sun \citep{Mayor1995}. Since then, astronomers have developed new techniques to detect and characterize exoplanets \citep[][]{Seager2010, Perryman2011}. The two more successful techniques are the transit and radial velocity (RV) methods \citep[see, e.g.,][]{Bozza2016}. They provide a direct measurement of the planet radius \citep{Charbonneau2000,Henry1999} and minimum mass \citep{Mayor1995}, respectively. By combining transit photometry with RV measurements, it is possible to measure the true mass and determine the planetary mean density. This allows us to to study planets' internal structure and composition and gives us important hints as to their formation and evolution.

The success of the transit method relies on both ground- \citep[e.g., \texttt{HAT-Net}, \texttt{KELT}, \texttt{WASP};][]{Bakos2004,Pepper2007,Pollacco2006} and space-based photometric surveys \citep[e.g., \texttt{CoRoT}, \texttt{Kepler}, and \texttt{K2};][]{Auvergne2009, Borucki2010, Howell2014}, which have led to the discovery of more than 2800 transiting exoplanets\footnote{As of July 2018, \url{exoplanet.eu}.}. The RV method has strongly benefited from state-of-the-art high-precision ($\sim$1\,\ms) spectrographs such as \texttt{HIRES} \citep{Vogt1994}, \texttt{HARPS} \citep{Mayor2003}, and \texttt{HARPS-N} \citep{Cosentino2012}, which have opened up the doors to the Earth-mass domain. Future exoplanet surveys and follow-up observations conducted with both space-based \citep[\texttt{TESS}, \texttt{PLATO}, \texttt{CHEOPS};][]{Ricker2015, Rauer2014, Broeg2013} and ground-based  \citep[e.g., \texttt{ESPRESSO}, \texttt{SPIRou};][]{Pepe2010,Donati2017} facilities will provide us with a wealth of photometric and spectroscopic time-series data sets that need powerful tools for a fast and robust analysis.

Radial velocity and transit light curves are described by time-dependent parametric equations. By comparing models with data we can determine the physical properties of planetary systems. Bayesian model fitting techniques, such as Markov chain Monte Carlo (MCMC) methods, are widely used for parameter estimation and their popularity among the astronomers' community has steadily increased in the past two decades \citep[see, e.g., ][]{Sharma2017}. 

Markov chain Monte Carlo data analysis is a reliable statistical method to estimate planetary parameters. However, it can be computationally challenging, especially when the amount of data is large and the dimension of the parameter space that one needs to explore is high. Some examples of such cases include: long-cadence photometric data -- as those collected by \texttt{Kepler} and \texttt{K2} -- for which the transit model has to be re-sampled to account for the long integration time \citep{Kipping2010}; the light curve of stars hosting ultra-short-period transiting planets $(P\,<\,1$ day), which includes hundreds or even thousands of transits \citep[e.g.,][]{Gandolfi2017,Guenther2017,Sanchis-Ojeda2014}; low-mass planets whose masses can be precisely derived only collecting hundreds of RV data points \citep[e.g.,][]{Hatzes2011}; multi-planet systems whose parameter space increases dimensionality with the addition of each planet \citep[e.g.,][]{Gillon2017}. Combining sophisticated statistical methods and powerful numerical tools is therefore an optimal approach to simultaneously model photometric and RV data.

There is a variety of software packages in the literature that allow us to determine the physical parameters of exoplanets from time-series photometry and RV measurements, using either purely Bayesian approaches \citep[e.g. \texttt{PYTRANSIT},\texttt{PyTranSpot};][]{Parviainen2015,Juvan2018}, or different methods \citep[e.g. \texttt{TLMC}][]{Csizmadia2015}. Some of these packages can model transit light curves (\texttt{PYTRANSIT}; \texttt{PyTranSpot}), RV curves \citep[e.g., \texttt{RVLIN},][]{Wright2009}, or even perform a joint analysis of the photometric and RV data \citep[e.g., \texttt{TLMC}; \texttt{EXOFAST},][]{Eastman2013}.

In this work we present the software suite \texttt{pyaneti}\footnote{From the Italian word \emph{pianeti}, which means \emph{planets}.}. This code is a new powerful tool to perform multi-planet fit to RV and/or transit data sets. It combines the MCMC technique with the computational power of \fortran\ and the versatility of \python.  This code has already been used for the analysis of several planetary systems \citep[see, e.g., ][]{Barragan2016,Barragan2018a,Barragan2018b,Chakraborty2018,Fridlund,Gandolfi2017,Guenther2017,Li2017,Livingston2018}.

The paper is organized as follows. For the sake of self-consistency, we provide a short recap of Bayesian analyses and MCMC algorithms in Sect.~\ref{sec:bayesian}. The RV and transit equations used by \texttt{pyaneti} are given in Sect.\,\ref{sec:equations}. Section~\ref{sec:code} describes the general algorithms used by the code. We test the package in Sect.\,\ref{sec:test} and conclude in Sect.\,\ref{sec:conclusions}.

\section{Mathematical approach}
\label{sec:bayesian}

\subsection{The Bayes' theorem}

The aim of data analysis is to extract information from experiments and/or observations. In this work, we are interested in extracting planetary physical parameters by comparing parametric models with astronomical observations. From a probabilistic point of view, we want to estimate the probability that a physical parametric model $M = M(\vec{\phi})$, function of some parameters $\vec{\phi}$, describes the data $D$. Such probability is called the conditional probability of $M$ given $D$ and it can be written as $P(M|D)$. 

Bayes' theorem \citep{Bayes} provides a simple and robust mathematical framework to compute $P(M|D)$ as

\begin{equation}
\label{eq:bayes}
P(M|D) = \frac{P(D|M)P(M)}{P(D)}
.
\end{equation}

In a context of a fixed dataset, $P(D|M)$ is a function of the model called the likelihood of observing the data set $D$ if the model $M$ is true, while $P(M)$ is the prior probability associated to the model $M$, and $P(D)$ is the model evidence. $P(M|D)$ is called the joint posterior distribution and it gives the probability that a model $M$ is true given $D$ is true. 
\subsection{Likelihood}

For a given data set $D$ composed of $N$ 
measurements $D_{1,\dots,N}$, we can generate a set of $N$ predicted points $M_{1,\dots,N}$ from a parametric model. The likelihood of a point $D_i$ being described by a point $M_i$ is written as $P(D_i|M_i)$. The likelihood of the whole data set $D$ to be described by the model $M$ is given by the product of each probability $P(D_i|M_i)$ as  

\begin{equation}
P({D}|{M}) =
\prod_{i=1}^N P(D_i|M_i).
\label{eq:multlike}
\end{equation}

In order to compute $P(D|M)$ using eq. (\ref{eq:multlike}), we need to find out which likelihood functions describes better our data. 
If we assume that our measurements are normally distributed, independent, and that only contain uncorrelated noise $\sigma_i$, the likelihood of the data point $D_i$ being true, assuming $M_i$ is also true, is written as

\begin{equation}
P(D_i|M_i) = \frac{1}{\sqrt[]{ 2\pi (\sigma^2_i + \sigma^2_{\rm j})} } \exp \left\{  
- \frac{1}{2}
\frac{ (D_i - M_i)^2 } {\sigma_i^2 + \sigma_{\rm j}^2}
\right\},
\label{eq:likelihooddatum}
\end{equation}

where the terms $\sigma_{\rm j}$ are used to normalize the likelihood in the case the nominal uncertainties $\sigma_i$ are underestimated \citep[see e.g., ][]{Sharma2017}. If we use eq. (\ref{eq:likelihooddatum}) for a data set, its likelihood is given by using eq. (\ref{eq:multlike}) as

\begin{equation}
P(D_i|M_i) = \prod_i^N \left[ \frac{1}{\sqrt[]{ 2\pi (\sigma^2_i + \sigma^2_{\rm j})} } \right] \exp \left\{ - \frac{1}{2} \chi^2 \right\}
,
\label{eq:likelihood}
\end{equation}

where 

\begin{equation}
\chi^2 = \sum_{i=1}^N \frac{ (D_i - M_i)^2 } {\sigma_i^2 + \sigma_{\rm j}^2}.
\label{eq:chi2}
\end{equation}
 
\texttt{pyaneti} uses the likelihood given by eq. (\ref{eq:likelihood}), but we acknowledge that more general likelihoods exist, in which possible correlated noise between data points is taken into account \citep[see, e.g., ][]{Parviainen2017,Sharma2017}. 
We also note that for numerical reasons, it is better to treat $P(M|D)$ in a logarithmic way (Appendix\,\ref{ap:ap1}).

\subsubsection{Priors}

Priors contain previously known information about a given model parameter, e.g., some physical range in which a parameter has equal probability to lie. Alternatively, the parameter's probability can be given by a distribution based on previous estimates. Widely used priors are the uniform and Gaussian priors. 

A \emph{uniform prior} is called a weakly informative or uninformative prior. It is used when the only available information about a given parameter $\phi_i$ is that it lies inside a range $[a,b]$. For example, we know that the eccentricity of an elliptical orbit ranges between $0$ and $1$. If the parameter $\phi_i$ lies between $a$ and $b$ with equal probability, its uniform prior is given by

\begin{equation}
\mathcal{U}(\phi_i;a,b) = \left\lbrace
     \begin{array}{cl}
       (b-a)^{-1} & : a < \phi_i < b \\
       0 & : {\rm otherwise}
     \end{array}
\right
.
\end{equation}

A \emph{Gaussian prior} is called an informative prior. This prior is useful when, for a given parameter, we have a previous measurement and its 1-$\sigma$ uncertainty, and we want to use this information to weight the probability. For instance, if we have asteroseismology-derived mass and radius of a star hosting a transiting planet, we can use these quantities together with the orbital period to set a Gaussian prior on the semi-major axis of the planet's orbit through Kepler\,third\,law. 

A Gaussian prior of a given parameter $\phi_i$ with median $a$ and standard deviation $b$ is given as

\begin{equation}
\mathcal{N}(\phi_i;a,b) = \frac{1}{\sqrt{2\pi b^2}} \exp \left[ - \frac{(\phi_i - a)^2}{2 b^2} \right].
\end{equation}

\begin{figure*}
  \center	
  \includegraphics[width=0.9\textwidth]{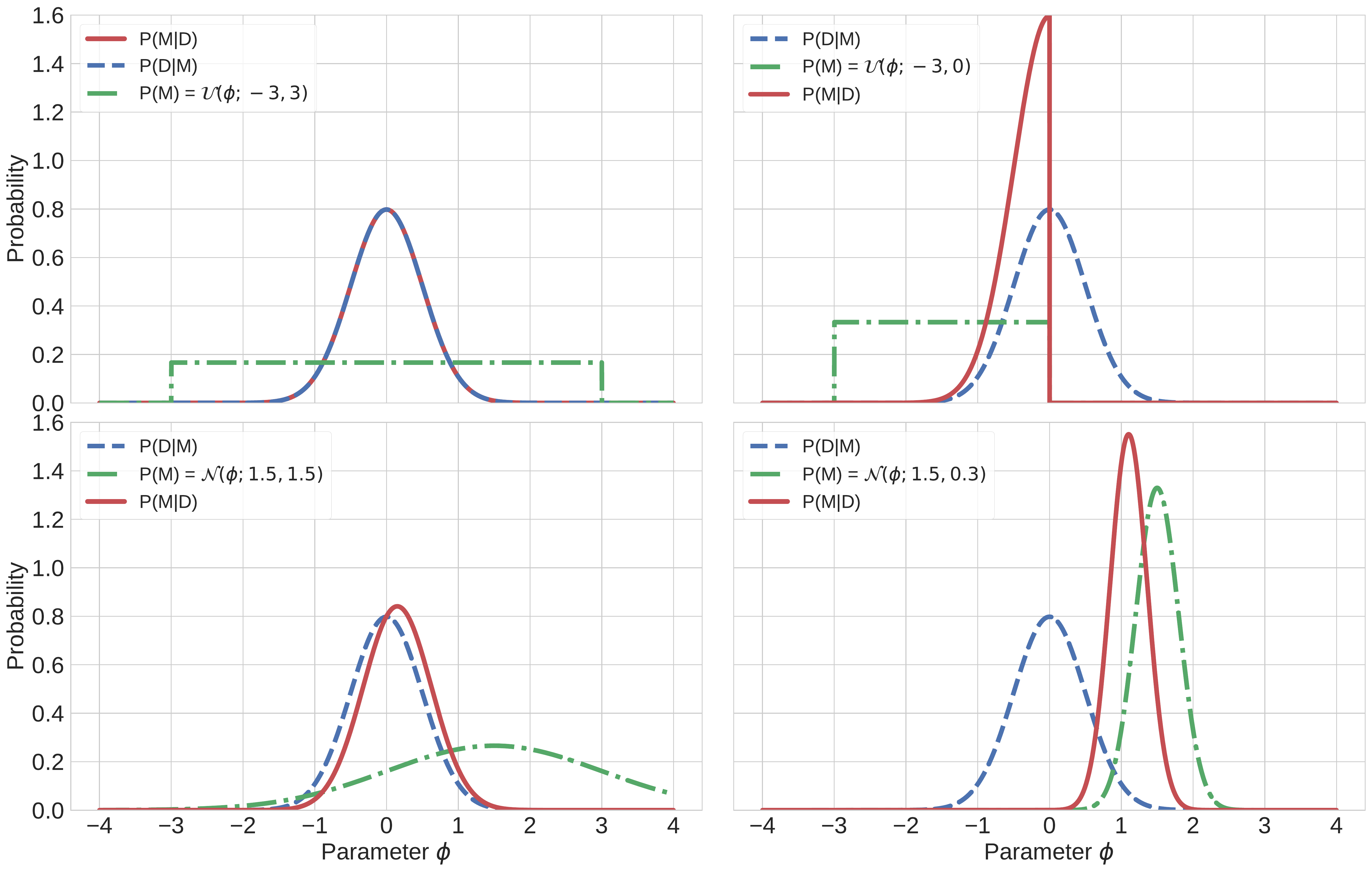}
  \caption{Posterior distributions (solid red line) for a fixed likelihood (blue dashed lines) and different priors (green dot-dashed lines). All quantities have been normalized for comparison purpose. \emph{Upper left:} Uniform prior with limits $[-3,3]$. \emph{Upper right:} Uniform prior with limits $[-3,0]$. \emph{Lower left:} Gaussian prior with mean 1.5 and standard deviation 1.5. \emph{Lower right:} Gaussian prior with mean 1.5 and standard deviation 0.3. \label{fig:priors}}
\end{figure*}

In this work we describe only the uniform and Gaussian priors, as those are currently implemented in \texttt{pyaneti}. We acknowledge the existence of other priors in the literature \citep[see, e.g., ][]{Diaz2018,Sharma2017}. Figure\,\ref{fig:priors} shows how priors can affect the final posterior distribution for a fixed likelihood. For instance, the upper and lower limits of a flat prior may truncate or exclude the maximum of the likelihood function. The influence of a Gaussian prior on the posterior distribution depends on the prior's center and width, as well as on the number of data points \citep[e.g., ][]{Gelman2004}.

\subsubsection{Model evidence}

The term $P(D)$ in equation (\ref{eq:bayes}) is called model evidence or marginal likelihood. It has the function to normalize the posterior distribution. By definition $P(D)$ is calculated by integrating the likelihood and prior distributions in the parameter space as 

\begin{equation}
P(D) = \int P(D|M(\vec{\phi}))P(M(\vec{\phi})) d \vec{\phi}.
\label{eq:evidence}
\end{equation}

\subsubsection{Marginal posterior distribution}

We now have a mathematical description of all the components to calculate $P(M|D)$ from equation (\ref{eq:bayes}). In order to derive the parameters, we are interested on the shape of the posterior distribution of each parameter $\phi_i$ more than in its normalized probability. The parameter estimation can be extracted from the non-normalized posterior distribution, i.e. the term $P(D|M)P(M)$, ignoring the evidence term $P(D)$. We note that the evidence term has an important role when doing Bayesian comparison between different models \citep[see, e.g., ][]{Gelman2004}.

Since $M$ is a parametric model, we can marginalize the parameter $\phi_i$ by integrating $P(D|M)P(M)$ over the remaining $\phi_{j \neq i}$ parameters. This leads to a marginal posterior distribution for each parameter $\phi_i$ from which we can infer the model parameters.

\subsection{Markov chain Monte Carlo}
\label{sec:mcmc}

The calculation of a marginal posterior distribution can be done analytically or numerically. However, in some cases it may not have an analytic solution. For instance, numerical iterative methods are widely used to sample the parameter space in order to generate marginal posterior distributions from a collection of data points.

An efficient method to generate a set of data points in a parameter space is by using a \emph{Markov chain}. Following the definition of \citet{Sharma2017}, a Markov chain is a sequence of random variables $X_1,\dots, X_n$ such that, given the present state, the future and past are independent. If random numbers are used to generate the Markov chains, this method is called Markov chain Monte Carlo (MCMC). These random variables can be the points $\vec{\phi}$ in the parameter space that we want to sample. For instance, if we start a point in the parameter space $\vec{\phi}_1$, we can generate a set of different models $\vec{\phi}_i$ via Markov chains. In this way, we can create a set of $N$ models from an initial $\vec{\phi}_1$.   

There is a large variety of MCMC sampling methods, which ensure that the Markov chains converge to the optimal solution where the posterior has a static solution. For a basic MCMC algorithm, we refer the reader to the Metropolis-Hastings algorithm \citep{Metropolis1953,Hastings1970}. In the next section we will describe the ensemble sampler algorithm \citep{Goodman2010} that is used by \texttt{pyaneti} for the parameter estimations. This algorithm was first used by \citet{Hou2012} to infer parameters from time-series RV measurements.
 
\subsection{Ensemble sampler algorithm}
\label{sec:ensemble}

The \emph{ensemble sampler} algorithm uses a group of Markov chains to explore the parameter space. 
Each chain $j$ starts with a point in the parameter space $\vec{\phi}_{j,t}$ and is evolved using the complementary chains of the ensemble.

\citet{Christen2006} found that it is possible to evolve the chain $\phi_{j,t}$ to the state $t+1$ via a {\it walk move} using a complementary chain of the ensemble. \citet{Goodman2010} used the idea of the walk move to construct an affine invariant move called {\it stretch move}. The stretch move for the chain $\vec{\phi}_{j,t}$ is defined as

\begin{equation}
\vec{\Phi}_j = \vec{\phi}_{k,t} + z \left( \vec{\phi}_{j,t} - \vec{\phi}_{k,t} \right),
\end{equation}

where $\vec{\phi}_{k,t}$ is a complementary chain of the ensemble, such that $j \neq k$ and $z$ is a scaling variable that regulates the step. This scaling variable has to come from a density distribution $g$ with the symmetry condition \citep{Christen2006}

\begin{equation}
g \left( \frac{1}{z} \right) = z\,g(z).
\label{eq:gden}
\end{equation}

A distribution that follows this condition is

\begin{equation}
   g(z) \propto \left\{
     \begin{array}{lr}
       \frac{1}{\sqrt{z}} & : z \in  \left[ \frac{1}{a}, a \right] \\
       0 & : {\rm otherwise},
     \end{array}
   \right.
\end{equation}

where $a\,>\,1$. There is no optimal value for $a$, but we set $a\,=\,2$ to be consistent with ensemble sampler algorithms in the literature \citep[e.g., ][]{Goodman2010,Hou2012}. To ensure the invariant distribution we have to compute the ratio

\begin{equation}
q = z^{N-1} \frac{P(M(\vec{\Phi})|D)}
{P(M(\vec{\phi}_{j,t})|D)}.
\label{eq:qensemble}
\end{equation}

The term $z^{N-1}$ ensures detailed balance \citep[for more details see][]{Goodman2010}. To decide whether we accept or not the proposed state we use

\begin{equation}
     \begin{array}{lr}
       \vec{\phi}_{j,t + 1} = \vec{\Phi}_j & : q > U \\
       \vec{\phi}_{j,t + 1} = \vec{\phi}_{j,t} & : q \leq U,
     \end{array}
     \label{eq:upchain}
\end{equation}

where $U$ is a random number between $[0,1]$. After a number $N$ of iterations and $L$ chains, we will have $N \times L$ samples for each parameter from which we can create posterior distributions. A general overview of a single step of the ensemble sampler method is given in Algorithm~\ref{al:ensemble}.

\begin{figure*}
  \center	
  \includegraphics[width=0.9\textwidth]{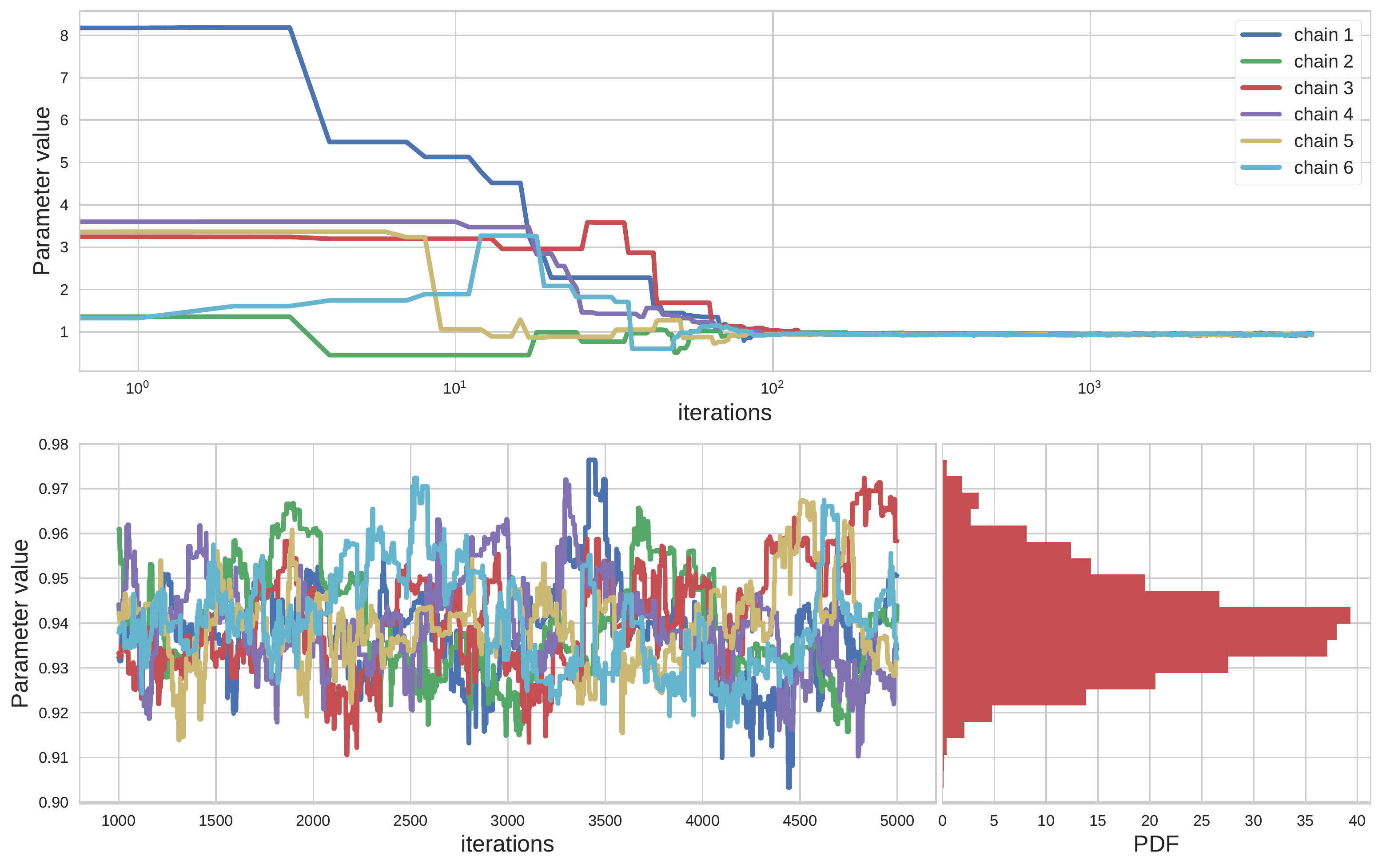}
  \caption{Example of the evolution of an ensemble sampler algorithm using six chains. \emph{Upper panel:} Parameter value for each chain from iteration 0 to 5000 in logarithmic scale. \emph{Lower left panel}: Chains behavior for the last 4000 iterations. \emph{Lower right panel:} Histogram created using the information contained in all the chains in the last 4000 iterations. \label{fig:chains}}
\end{figure*}

Figure \ref{fig:chains} shows an example of the evolution of an ensemble sampler algorithm using six chains. The latter start at a different point in the parameter space. After a finite number of iterations (in this case a few hundreds), the chains converges to a stable region of the parameter space. Details on how we create marginal posterior distributions from chain's samples are provided in Section\,\ref{sec:marginalchains}.

Another advantage of this approach is that, since each Markov chain evolves independently, this algorithm can be parallelized \citep{Foreman2013}.

\IncMargin{1em}
\begin{algorithm}
\SetKwData{Left}{left}\SetKwData{This}{this}\SetKwData{Up}{up}
\SetKwFunction{Union}{Union}\SetKwFunction{FindCompress}{FindCompress}
\SetKwInOut{Input}{input}\SetKwInOut{Output}{output}
\Input{Initial ensemble of $N$ states $\vec{\phi}_{j,t}$}
\Output{Ensemble of $N$ states $\vec{\phi}_{j,t+1}$}
\BlankLine
\For{$j = 1$ \KwTo $N$}{
Select a complementary state from the ensemble such that $j \neq k$\\
Sample the scaled variable $z$ from the density distribution $g$\\
Propose the new state via a walk move 
$\vec{\Phi}_j = \vec{\phi}_{k,t} + z \left( \vec{\phi}_{j,t} - \vec{\phi}_{k,t} \right)$\\
Compute $q$ from eq. (\ref{eq:qensemble}) using likelihood and priors for the states $\vec{\phi}_{j,t}$ and $\vec{\Phi}_j$\\
Sample an uniform random variable $U$ between 0 and 1\\
\eIf{$q > U$}{
$\vec{\phi}_{j,t+1} = \vec{\Phi}_j$\\
}{
$\vec{\phi}_{j,t+1} = \vec{\phi}_{j,t}$\\
}
}
\caption{One iterations of the ensemble sampler algorithm. \label{al:ensemble}}
\end{algorithm}\DecMargin{1em}

\subsection{Convergence}

In order to infer the parameter values based on a MCMC sampling we need to use chains that have converged. A widely used convergence test has been developed by \citet{Gelman1992}. This test compares the ``between-chain'' $B$ and ``within-chain'' $W$ variance via the scaled potential factor $\hat{R} = \sqrt{[W(n-1)/n + B/n]/W}$, where $n$ is the length of each chain. We define convergence as when chains have $\hat{R} < 1.02$ for all the parameters \citep{Gelman2004}.

\subsection{Marginal posterior distribution from parameter sampling}
\label{sec:marginalchains}

Chains that have converged to a static solution represent a sample of the marginal posterior distribution from which they were sampled. The frequencies of the chains can be used to create the posterior distribution of the sampled parameters. A common way to draw the sampling frequency is with a histogram, as shown in Fig,\,\ref{fig:chains}. The final marginal posterior distribution for each parameter is also called \emph{credible interval}. 

The median and the 68\% limits of the credible interval are commonly used to define the parameter's best estimate and its uncertainty \citep[see, e.g., ][]{Hogg2018}. When the marginal posterior distribution follows a Gaussian distribution, the median and the 68\% limits of the credible interval correspond to the mean and standard deviation of a normal distribution. When the posterior distribution is skewed, the 68\% limits are not symmetric with respect to the median, and they give an ``first-order'' idea of the shape of the marginal posterior distribution that describes a given parameter.

\section{Multi-planet equations}
\label{sec:equations}

In this section, we describe the equations used by \texttt{pyaneti} to extract parameter values from time-series RV and photometric transit data. For a detailed derivation of the equations presented in the following sub-sections, we refer the reader to the specific literature \citep[see, e.g.,][]{Murray2010,2010exop.book...55W}.

\subsection{Radial velocity equations}
\label{RV_Equations}

The motion of a planet orbiting a star can be reduced to a two-body problem in which one of the two body is far more massive than the other. The change of the velocity component along the line-of-sight to the host star, induced by the presence of the orbiting planet, is described by the following equation

\begin{equation}
f_{\rm \,RV}(\vec{\phi};t) = \gamma + 
K_\star \left[ 
\cos \left( \theta + \omega_{\star} \right) + e \cos\omega_{\star} 
\right],
\label{eq:rvplanet}
\end{equation}

where $\gamma$ is the systemic velocity of the center of mass, $\theta$ is the true anomaly, $e$ the orbital eccentricity, and $\omega_{\star}$ the angle of periastron of the star. $K_\star$ is the radial velocity semi-amplitude variation, which is given by 

\begin{equation}
K_\star = \left( \frac{2 \pi {\rm G}}{P} \right)^{1/3} 
\frac{ M_{\rm p} \sin i}{  \left( M_{\rm p} + M_\star \right)^{2/3} }
\frac{1}{ \left( 1 - e^2 \right)^{1/2}},
\label{eq:massf}
\end{equation}

Equation (\ref{eq:massf}) provides a relation between the semi-amplitude RV variation of the star $K_\star$ and the planetary $M_{\rm p}$ and stellar mass $M_\star$. The remaining parameters of eq. (\ref{eq:massf}) are the gravitational constant ${\rm G}$ \citep[we use the IAU units given by][]{Prsa2016}, the orbit inclination with respect to the line-of-sight $i$, and the orbital period $P$. Since $M_{\rm p} << M_\star$, we can assume that $(M_{\rm p} + M_\star)^{2/3} \approx M_\star^{2/3}$. The dependence on the orbit inclination implies that only the planet's minimum mass $M_{\rm p} \sin i$ can be measured, provided that the stellar mass $M_\star$ is known. In order to determine the planet's true mass, the orbit inclination has to be measured. For transiting exoplanets, the inclination can be derived from the modeling of the transit light curve (Sect.\,\ref{sec:transits}).

The time dependency of $\theta$ is given by 

\begin{equation}
\theta(t) = 2 \arctan \left[ 
\sqrt[]{\frac{1 + e}{ 1- e }}
\tan \left( \frac{E(t)}{2} \right)
\right]
\label{eq:true}
\end{equation}

where the eccentric anomaly $E(t)$ is defined via

\begin{equation}
A(t) = E(t) - e \sin[E(t)]
\label{eq:ecc}
\end{equation}

and where

\begin{equation}
A(t) = \frac{2 \pi}{P} \left(  t - T_{\rm p} \right)
\label{eq:mean}
\end{equation}

is the mean anomaly. The latter depends on time $t$ and on a zero point $T_{\rm p}$, which is the time of passage of periastron. We stress that we fit for the time of minimum conjunction of the planet $T_0$ that, for transiting planets, coincides with the time of transit . It is straightforward to pass from $T_0$ and $T_{\rm p}$ since the corresponding points in the planet's orbit are separated by an angle of $\pi/2 - \omega_\star$. 
We note that eq.\,(\ref{eq:ecc}) is transcendent and cannot be solved analytically with respect to the eccentric anomaly $E(t)$. \texttt{pyaneti} solves eq.\,(\ref{eq:ecc}) using a Newton-Raphson algorithm.

If a star is orbited by $N_{\rm p}$ planets -- which we assume their mutual gravitational interaction is negligible -- the general expression for eq.\,(\ref{eq:rvplanet}) is then

\begin{equation}
M_{\rm \, RV}(\vec{\phi};t) = \gamma_i + \sum_{j=1}^{N_{\rm p}}
K_j \left[ 
\cos(\theta_j + \omega_{\star,j}) + e_j \cos\omega_{\star,j}
\right].
\label{eq:rvmulti}
\end{equation}

The term $\gamma_i$ depends on the spectrograph $i$ to account for possible instrumental offsets. Equation\,(\ref{eq:rvmulti}) can be modified to add linear or quadratic acceleration terms that may be present in the data.

The general parametric function $M_{\rm \, RV}(\vec{\phi};t)$, which describes the Doppler reflex motion of a star orbited by more than one planet is then

\begin{equation}
M_{\rm RV}(\vec{\phi};t) = f(\{ T_{0},P,e,w_{\star},K \}_j,\gamma_i;t),
\label{eq:genrv}
\end{equation}

in which the set of parameters $\{T_{0},P,e,w_{\star},K\}$ repeats for each planet $j$ and $\gamma_i$ accounts for each different instrument\,$i$.

\subsection{Transit equations}
\label{sec:transits}

An eclipse occurs when an astronomical body is obscured by a second one. A transit is a special case of eclipse, in which a smaller object passes in front of a larger body. If the orbit inclination is close to $90^\circ$, the presence of a planet orbiting its host star can be inferred by detecting the periodic drops of stellar flux caused by the planet partly occulting the stellar disk. The fraction of light occulted by a planet during a transit is proportional to its size, being about 1\,\% for a Jupiter-size object and 100 times smaller for an Earth-size planet transiting a Solar-like star.

An useful quantity to describe a planetary transit is the scaled projected distance between the planet's and the star's center defined as \citep[see, e.g.,][]{2010exop.book...55W}

\begin{equation}
	\delta = \frac{a}{R_\star} \frac{\left( 1 - e^2\right)}{ \left( 1 + e \cos \theta(t) \right) }
	\sqrt{1 - \sin^2\left( \theta(t) + \omega_{\star} \right) \sin^2 i},
     \label{eq:distsky}
\end{equation}

where $a$ is the semi-major axis of the relative orbit\footnote{The semi-major axis of the relative orbit is defined as $a=a_p+a_\star$, where $a_p$ and $a_\star$ are respectively the semi-major axes of the planet's and star's orbit with respect to the center of mass.}, $R_\star$ is the stellar radius, and the remaining parameters are the same used in the RV equations presented in Sect.\,\ref{RV_Equations}. We note that $\delta$ depends on the true anomaly $\theta(t)$, which in turn is a function of time according to eqs.\,(\ref{eq:true}), (\ref{eq:ecc}) and (\ref{eq:mean}). Following \citet{Eastman2013}, we define the projected distance $\delta$ using the argument of periastron of the star $\omega_\star$ instead of the argument of periastron of the planet $\omega_p$. If we define $r_{\rm p} \equiv R_{\rm p}/R_\star$ as the planet-to-star radius ratio, from equation (\ref{eq:distsky}), the transit of an exoplanet occurs only when $\delta < 1 + r_{\rm p} $ and $\sin\,(\theta(t) + \omega_\star)\,>\,0$ (star behind the planet). On the other hand, the planet's occultation -- also known as secondary eclipse -- occurs if $\delta < 1 + r_{\rm p} $ and $\sin\,(\theta(t) + \omega_\star) < 0$ (planet behind the star).

In order to analytically describe how the total flux $F(t)$ changes as a function of time due to the presence of a transiting planet, we need to account for the disk-integrated stellar flux $F_\star (t)$, the planet flux $F_{\rm p}(t)$ (both reflected light and thermal emission), and the loss of light when transits/occulations occur $\lambda (\delta,r_{\rm p})$. The total flux $F(t)$ is\,given\,by 

\begin{equation}
F(t) = F_\star(t) + F_{\rm p}(t) - \lambda (\delta,r_{\rm p}).
\label{eq:totalflux}
\end{equation}

We assume that the stellar flux $F_\star(t)$ is constant and equal to 1, and that any variation can be expressed as a fraction of the stellar light. We also assume that the planet contribution to the light curve is negligible ($F_{\rm p} = 0$), i.e., we assume that occulations and phase curve have no effect on the observed light curve. Under these assumptions, eq.\,(\ref{eq:totalflux}) can be re-written\,as

\begin{equation}
F(t) = 1 - \lambda (\delta,r_{\rm p}).
\label{eq:lostofflux}
\end{equation} 

By definition $\lambda = 0$ when $\delta > 1 + r_{\rm p}$. For cases where $\delta < 1 + r_{\rm p}$, the change of light depends on the analytical form of $\lambda$, which accounts for the loss of light as the planet crosses the stellar disk. There are different approaches to define the analytical form of $\lambda (\delta,r_{\rm p})$. The simplest approach is to assume that the stellar disk is a uniform source of light \citep[][]{Seager2003}. However, real stellar disks are brighter in the center and fainter at the edge (the limb), a phenomenon known as limb darkening \cite[see, e.g.,][]{Claret2011}. \texttt{pyaneti} uses the \citet{2002ApJ...580L.171M}'s transit light cure model in which the stellar intensity is limb-darkened using a quadratic law with coefficients $u_1$ and $u_2$. \citet{2002ApJ...580L.171M} provide the function $\lambda (\delta,r_{\rm p})$ for a single planet transiting a star. 

For a system where there are $N_{\rm p}$ transiting planets, the relative flux of the star is then 

\begin{equation}
	F(t) = 1 - \sum_{j=1}^{N_{\rm p}} \lambda_j(\delta, r_{\rm p}).
	\label{eq:multitransit} 
\end{equation}

According to eq.~(\ref{eq:multitransit}), $F(t) = 1$ if no planet transits the star; it reduces to the one-transiting-planet case when there is a single planet crossing the stellar disk. Equation\,(\ref{eq:multitransit}) takes also into account multi-planet transit events. We note that this approach does not take into account occultations between planets that may occur. 

The general parametric function $M_{\rm tr}(\vec{\phi})$ that describes transit events in a light curve is 

\begin{equation}
M_{\rm tr}(\vec{\phi};t) = f(\{ T_{0},P,e,w_{\star},
R_{{\rm p}}/R_\star,a/R_\star,i \}_j,\{u_1,u_2\}_i;t).
\label{eq:gentr}
\end{equation}

The set of parameters $\{ T_{0},P,e,w_{\star}, R_{{\rm p}}/R_\star,a/R_\star,i \}$ repeats for each planet $j$.
Each $\{u_1,u_2\}$ repeats for each band $i$ of the light curve. 

\subsection{Multi-planet joint fit}

When both Doppler and transit data are available, the best approach  to perform the analysis is via a joint fit. By comparing the RV and the transit equations (eqs.\,\ref{eq:genrv} and \ref{eq:gentr}), we see that $T_0$, $P$, $e$ and $\omega_\star$ are shared parameters. In this case the orbital parameters are fitted simultaneously. Usually, the transit light curves improve the determination of  $T_0$ and $P$, whereas the RV measurements constrain better $e$ and $\omega_\star$. The parameters involved in the joint fit are

\begin{equation}
\vec{\phi} = (\{ T_{0},P,e,w_{\star},
R_{{\rm p}}/R_\star,a/R_\star,i,K \}_j,\{u_1,u_2\}_i,\gamma_k),
\label{eq:jointfit}
\end{equation}

where $\{ T_{0}\,,P\,,e\,,w_{\star}\,,R_{{\rm p}}/R_\star\,,a/R_\star\,,i\,,K_{\star}\}_j$ repeats for each planet $j$, $ \{u_1,u_2\}_i $ for each photometric band $i$, and $ \gamma_K $ for each spectrograph $k$.

\subsection{Parametrizations}
\label{sec:parametrization}

Equation (\ref{eq:jointfit}) defines the general set of parameters that can be extracted by modeling RV measurements and transit photometry. It is possible to use a set of convenient parametrizations to improve the exploration of the parameter space and avoid biases due to priors. In the following sub-sections we provide a brief description of the parameterizations used by \texttt{pyaneti}.

\subsubsection{Eccentricity and angle of periastron}

The posterior distribution of the eccentricity is not well sampled for orbits with small eccentricities \citep{Lucy1971}. A practical solution is to define $e$ and $\omega_\star$ using a polar form. \texttt{pyanety} adopts the parametrization proposed by \citet{Anderson2011}

\begin{equation}
ew_1 = \sqrt{e} \sin \omega_\star, \hspace{0.5cm}
ew_2 = \sqrt{e} \cos \omega_\star.
\end{equation}

This parameterization has two advantages: $a$) it is not truncated when the eccentricity is close to zero; $b$) uniform priors on $ew_1$ and $ew_2$ imply uniform priors on the eccentricity. 

\subsubsection{Impact factor}

As presented in Sect.\,\ref{sec:transits}, the transit of a planet can described using the scaled projected distance between the planet and star centers. It is then convenient to parametrize the stellar inclination using a parameter that takes into account the projected distance. A practical approach is via the impact parameter defined as \citep{2010exop.book...55W}

\begin{equation}
b = \frac{a}{R_\star} \cos i
\left(
\frac{1-e^2}{1+e \sin \omega_\star}
\right).
\label{eq:impactf}
\end{equation}

The advantage of using the impact factor is that $b$ can be compared directly with the projected distance $z$. In this way it is easy to set priors to exclude orbits for which there are no transit, i.e., when $b>1+r_{\rm p}$.

\subsubsection{Limb Darkening coefficients}

For the limb darkening coefficients \texttt{pyaneti} uses the parameterization proposed by \citet{Kipping2013}, who showed that an optimal way to sample the parameter space for the \citet{2002ApJ...580L.171M}'s limb darkening coefficients is via the parametrization 

\begin{equation}
q_1 = ( u_1 + u_2)^2, \hspace{0.5cm} q_2 = \frac{u_1}{ 2(u_1+u_2)}.
\end{equation}

The advantage of this approach is that it fully accounts for our ignorance about the intensity profile and explores physical solutions by sampling uniformly $q_1$ and $q_2$ between 0 and 1. This yields robust and realistic uncertainty estimates. It is possible to recover the original $u_1$ and $u_2$ coefficients via

\begin{equation}
u_1 = 2 q_1 \sqrt{q_2} , \hspace{0.5cm} u_2 = \sqrt{q_1} ( 1 - 2 q_2  ).
\end{equation}

\subsubsection{Stellar density}

From Kepler's third law we obtain that

\begin{equation}
\rho_\star + r_{\rm p}^3 \rho_{\rm p} = \frac{3 \pi}{{\rm G} P^2}
\left(
\frac{a}{R_\star}
\right)^3.
\label{eq:dena}
\end{equation}

where $\rho_\star$ is the star's mean density, $\rho_p$ the planet's mean density, $r_{\rm p}$ the planet-to-star radius ratio, $P$ the orbital period, $R_\star$ the star's radius, and $a$ the semi-major axis of the relative orbit. Since $r_{\rm p}^3$ is relatively small, the second term of the left side of eq.~(\ref{eq:dena}) can be neglected \citep{2010exop.book...55W}. There is thus a relation between the stellar density and the orbital parameters $P$ and $a/R_\star$ that can be used to compare stellar density derived from the modeling of the transit light curves with an independent determination (e.g., from spectroscopy). 

It is convenient to parametrize $a/R_\star$ with $\rho_\star$. If precise stellar parameters have been calculated, it is possible to set tight priors on the stellar density and hence on $a/R_\star$. For a multi-planet system, it is convenient to parameterize the scaled semi-major axis $a_j/R_{\star}$ of all planets $j$ using the same stellar density. In this way the stellar density is constrained for all planets and Kepler's third law is not violated within planets orbiting the same star.  

\texttt{pyaneti} uses the parametrization $\rho_\star^{1/3}$ instead of $\rho_\star$ because $a/R_\star$ and $\rho_\star^{1/3}$ are linearly related assuming $r_{\rm p}\,\approx\,0$ (eq.\,\ref{eq:dena}). 

\section{Code overview}
\label{sec:code}

If we combine the MCMC analysis described in Sect.\,\ref{sec:bayesian} along with the multi-planet equations presented in Sect.\,\ref{sec:equations}, we can develop a powerful tool to estimate planet parameters from Doppler and transit observations. We used this approach together with the computational speed of \fortran\ and the versatility of \python\ to write the software suite \texttt{pyaneti}. The computation-demanding routines, such as orbital solutions, likelihood calculations, etc., are calculated by \fortran\ subroutines. The input and output routines, such as data preparation, plot creations, etc., are handled by \python. \fortran\ subroutines are wrapped to \python\ using \texttt{F2PY}\footnote{More documentations are available at \url{http://www.f2py.com/}.}.

One of the main advantages of \texttt{pyaneti} is that all the code controls are given inside a \python-based input file. Priors, fitted parameters, and data files are controlled via flags and python objects. This allows one to run the code with only one command line. A general overview of the algorithm of the code is given in Algorithm \ref{al:pyaneti}.

\IncMargin{1em}
\begin{algorithm*}
\SetKwInOut{Input}{input}\SetKwInOut{Output}{output}
\Input{RV and/or light curve time-series, correct input file for \texttt{star-name}}
\Output{Posterior distributions, plots, parameter inference of RV and/or transit models from data}
\BlankLine
{
{\texttt{./pyneti.py star-name} (start of the run)}\\
Read initial files (functions, default values)\\
Read input file with parameters for the current run (number of planets, priors, flags) \\
Read time-series data\\
Pass data and variables to \fortran\ routines\\
}
{
Start \fortran\ execution \\
Create random chains inside the prior ranges\\
Calculate likelihoods and priors for the initial state\\
Set iteration control variable \texttt{continua} to \texttt{True} \\
Initialize iterations count variable \texttt{i=0}\\
\While{ \texttt{continua} }{
Evolve chains following Algorithm \ref{al:ensemble} \hfill ! This line can also run in parallel\\
check for convergence after $N$ iterations \hfill !$N$ is calculated as \texttt{niter} $\times$ \texttt{thin\_factor} \\
  \If{ \texttt{i == $N$} }{
    Check convergence using \citet{Gelman1992} criteria \\
    \eIf{ \texttt{chains converged == False}  }{  
    \texttt{continua = True }    \hfill !Chains have not converged: Keep iterating\\
    i = 0 \hfill ! restart iteration counter \\
    }{
    \texttt{continua = False }  \hfill !Chains have converged: save posteriors\\
    }
  }
}
Write posterior distributions with the converged chains taking into account the thin factor\\
End of \fortran\ execution\\
}
{
Run \python\ output routines \\
Read posterior from posterior file\\
Automatic calculation of parameters and creation of plots\\
Save data in the \texttt{outpy/star-name\_out} directory\\
End of run for \texttt{star-name}\\
}
\caption{General algorithm of \texttt{pyaneti}.  \label{al:pyaneti}}
\end{algorithm*}\DecMargin{1em}

The advantage of using the ensemble sampler algorithm described in section \ref{sec:ensemble} is that it can be parallelized. This speed-up the global solution of the MCMC run. The parallelization is done following the procedure described in \citet{Foreman2013}, in which we divide the ensemble in two subsamples and evolve each group taking a chain from the complementary set of chains. We use Open Multi-Processing (\texttt{OpenMP}) to perform the parallelization inside the \fortran\ routines.

There are some physical effects that are not included in the current version of the code. Transit timing variations (TTVs), mutual interaction between planets, multi-band photometry, Rossiter-McLaughlin effect, planet's occultations, planet's phase curve fitting have not been implemented yet. \texttt{pyaneti} currently uses likelihood and priors as described in Sect.\,\ref{sec:bayesian}. More general likelihoods, such as Gaussian Process, have not been included yet. Nevertheless, the code is written in a modular way making it easy and straightforward to implement additional physical effects or equations. We plan to keep maintaining and upgrading \texttt{pyaneti}.

\section{Code tests}
\label{sec:test}

\subsection{A toy model}
\label{Synthetic_data}

\subsubsection{Setup}

We created a set of synthetic RV and transit data to check the performance of \texttt{pyaneti}. The simulated planetary system includes three planets: the two innermost planets transit the star, whereas the outer planet can only be seen in the RV data set. 

Synthetic data points were created assuming a star with a mass of 0.66\,$M_{\odot}$ and radius of 0.67\,$R_{\odot}$. The planets have periods of 1.21321, 5.61122, and 12.12349 days with conjunction times of 1.0, 2.21529, and 4.63963 days, respectively. Their radii and masses are 1.5, 3.0, and 7 $R_{\oplus}$, and 5, 10, and 62 $M_{\oplus}$, respectively. The orbits of the two innermost planets are circular, whereas the outer planet has a non-zero eccentricity of $e_{\rm c}=0.1$ with the star's argoment of periastron $\omega_{\star,{\rm c}} = 204 \deg$. We assumed inclinations of $i_{\rm b} = 87\,\deg$, $i_{\rm c} = 88\,\deg$, and $i_{\rm d} = 84\,\deg$, so that the two innermost planets transit the star while planet d does not. We assumed that gravitational interaction between the three planets is negligible. We imposed limb darkening coefficients of $u_1 = 0.43$ and $u_2 = 0.31$. We used these values to calculate the scaled parameters used by \texttt{pyaneti}. Details of the whole set of fitted parameters are given in Table\,\ref{tab:parameters}.

\begin{figure*}
\centering
  \includegraphics[width=0.8\textwidth]{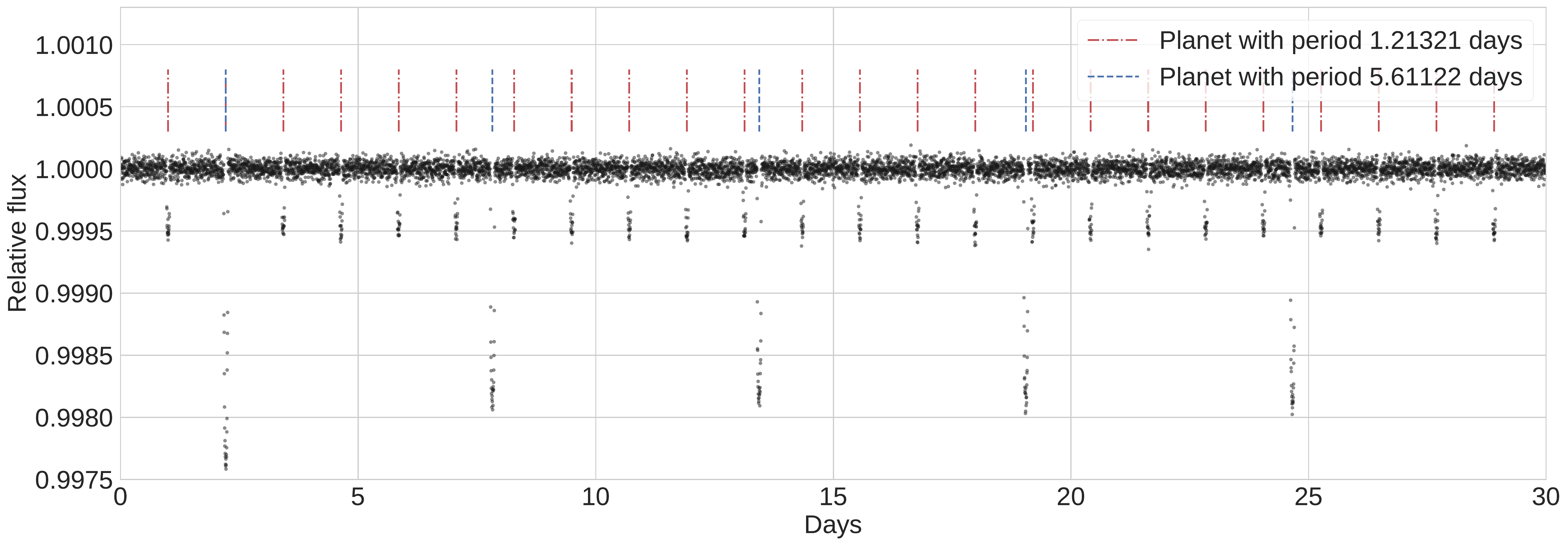} \\
  \includegraphics[width=0.8\textwidth]{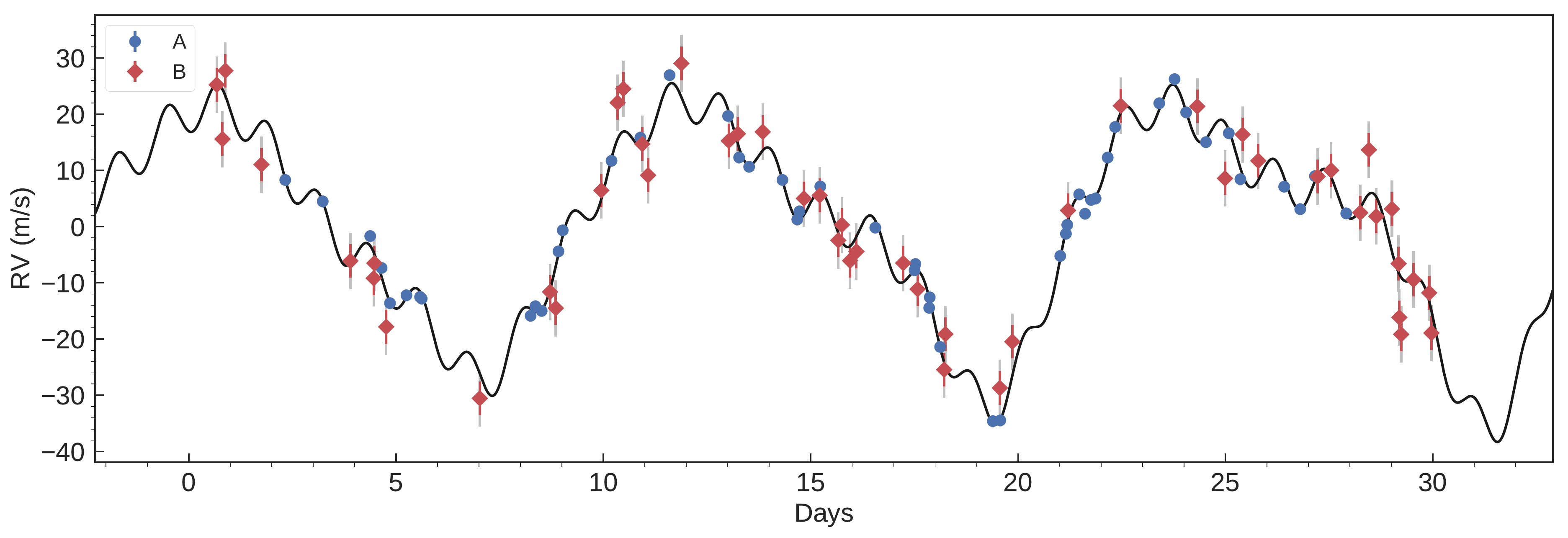} \\
  \caption{{\it Upper panel:} Thirty-day-long synthetic light curve. We assumed a relative flux of 1 with no trends. There are two transiting planets with different periods and sizes, marked with dashed lines. {\it Lower panel:} Synthetic RV measurements. The RV signal consists of three planets with different masses and periods. Instrument A (blue circles) has a precision of 1 \ms, whereas instrument B (red squares) has a precision of 5 \ms. The nominal error bars of instrument B were set to 3 \ms\ to simulate a jitter term (gray extensions to the nominal error bars). The best fitting model is shown as a thick black line. Offsets were subtracted to both data sets. 
\label{fig:syndata}}
\end{figure*}

The synthetic light curve covers a range of $30$ days starting at an arbitrary $0$ point. We created the instantaneous normalized flux due to the transiting planets using eq.~(\ref{eq:multitransit}) with continuous time stamps separated by 5 minutes. We added Gaussian noise at the $5\times10^{-5}$ level to simulate high precision photometry, such as that provided by \texttt{Kepler}. The synthetic light curve is displayed in the upper panel of Fig.\,\ref{fig:syndata}. 

The simulated RV measurements cover the 30 days simultaneous to the light curve data. Time stamps were taken from a random uniform distribution and the corresponding RVs were calculated using eq.~(\ref{eq:rvmulti}). We simulated data for two spectrographs called instrument A and B. For spectrograph A, we created 50 absolute RVs with Gaussian noise of 1\,\ms\ assuming a systemic velocity of 10\,\kms. For instrument B, we simulated 50 time stamps with Gaussian noise of 5\,\ms, but we assigned error bars of 3\,\ms\ to each point to simulate a jitter term of $\sigma_{\rm j} = (5^2-3^2)^{1/2} \,{\rm m\,s^{-1}} = 4$\ms. We assumed that the RVs of instrument B are relative and arbitrarily centered around 0\,\kms. The lower panel of Fig.~\ref{fig:syndata} shows the synthetic data-points following the correction for the RV offset between the two instruments.

We performed a joint fit setting uniform priors for all the parameters (details are given in Table\,\ref{tab:parameters}). We used 100 independent Markov chains started randomly inside the uniform prior ranges. Once all chains converged, we used the last 5000 iterations and saved the chains' states every 10 iterations. This approach generated a posterior distribution of 50,000 points for each parameter.

\subsubsection{Results}

Table \ref{tab:parameters} contains the medians and 68\% credible intervals of the posterior distributions of the fitted parameters. We note that the system parameter's true values are inside the posterior distribution of each parameter. In most cases, the true values are inside the 68\% credible interval. This shows the power of \texttt{pyaneti} to infer real parameters from data. 

\begin{table*}
  \caption{System parameters. \label{tab:parameters}}  
  \begin{tabular}{lccc}
  \hline
  Parameter & Real value & Prior$^{(\mathrm{a})}$ & Inferred value \\
  \hline
  \multicolumn{4}{l}{\emph{Model Parameters planet b}} \\
  \noalign{\smallskip}
    Orbital period $P_{\mathrm{orb}}$ (days) &1.21321 &  $\mathcal{U}[ 1.2122 , 1.2142 ]$ & $1.2132028 \pm 0.0000097  $ \\
    Transit epoch $T_0$ & 1.0 & $\mathcal{U}[ 0.9965 , 1.0035 ]$ & $1.00010 \pm 0.00012 $  \\  
    Scaled planet radius $R_\mathrm{p}/R_{\star}$ & 0.020525 & $\mathcal{U}[0,0.1]$ & $0.0204479 _{ - 0.00015} ^ { + 0.00027 } $  \\
    Impact parameter, $b$ & 0.33 & $\mathcal{U}[0,1]$  & $0.25_{ - 0.15 } ^ { + 0.13 } $ \\
    $e$ & 0 & $\mathcal{F}[0]$ & 0 \\
    $\omega_\star$ & $0$ & $\mathcal{F}[0]$ & 0 \\
    Radial velocity semi-amplitude variation $K$ (m s$^{-1}$) & 3.95 & $\mathcal{U}[0,50]$ &  $4.16 \pm 0.23 $ \\
  \hline
  \multicolumn{4}{l}{\emph{Model Parameters planet c}} \\
  \noalign{\smallskip}
    Orbital period $P_{\mathrm{orb}}$ (days) & 5.61122 &  $\mathcal{U}[ 5.6012 , 5.6212 ]$ & $5.611254 \pm 0.000041 $ \\
    Transit epoch $T_0$ & 2.21529 & $\mathcal{U}[ 2.2143 , 2.2163 ]$ & $2.21530 \pm 0.00010$  \\  
    Scaled planet radius $R_\mathrm{p}/R_{\star}$ & 0.04105 & $\mathcal{U}[0,0.1]$ & $0.04058 _{ - 0.00030 } ^ { + 0.00057 } $   \\
    Impact parameter, $b$ & 0.60 & $\mathcal{U}[0,1]$  & $0.57 _{ - 0.03 } ^ { + 0.06 } $ \\
    $e$ & 0 & $\mathcal{F}[0]$ & 0 \\
    $\omega_\star$ & $0$ & $\mathcal{F}[0]$ & 0 \\
    Radial velocity semi-amplitude variation $K$ (m s$^{-1}$) & 4.74 & $\mathcal{U}[0,50]$ & $4.86 \pm 0.33 $ \\
      \hline
  \multicolumn{4}{l}{\emph{Model Parameters planet d}} \\
  \noalign{\smallskip}
    Orbital period $P_{\mathrm{orb}}$ (days) & 12.12349 &  $\mathcal{U}[ 11.8235 , 12.4235 ]$ & $12.142 \pm 0.028 $ \\
    Transit epoch $T_0$ & 4.640 & $\mathcal{U}[ 4.1396 , 5.1396 ]$ & $4.592 \pm 0.052 $  \\  
    $e$ & $0.1$ & $\mathcal{U}[0,1]$ & $0.096 \pm 0.012 $ \\
    $\omega_\star$ & $3.57$ & $\mathcal{U}[0,2\pi]$ & $3.60 \pm 0.14$ \\
    Radial velocity semi-amplitude variation $K$ (m s$^{-1}$) & 22.75 & $\mathcal{U}[0,50]$ & $22.99 \pm 0.24$ \\
      \hline
      \multicolumn{4}{l}{\emph{Other Parameters}} \\
  \noalign{\smallskip}
      Cubic root of stellar density $\rho_{\star}^{1/3}$ & 1.458 &  $\mathcal{U}[0.05 , 2]$ & $1.496 _{ - 0.072 } ^ { + 0.039 } $ \\
      Systemic velocity $\gamma_{\rm A}$ (\kms) & 10 & $\mathcal{U}[9,11]$ & $9.99991 \pm 0.00017 $ \\
      Systemic velocity $\gamma_{\rm B}$ (\kms) & 0 & $\mathcal{U}[-1,1]$ & $0.00107 \pm 0.00084$ \\
      Jitter term $\sigma_{\rm A}$ (\ms) & 0  & $\mathcal{U}[0,100]$ & $0.25 _{ - 0.18 } ^ { + 0.24 } $ \\
      Jitter term $\sigma_{\rm B}$ (\ms) & 4 & $\mathcal{U}[0,100]$ & $4.09 _{ - 0.68 } ^ { + 0.77 }$ \\
    Parameterized limb-darkening coefficient $q_1$ & 0.55 & $\mathcal{U}[0,1]$ & $0.60 { \pm 0.05 } $  \\
    Parameterized limb-darkening coefficient $q_2$ & 0.29 & $\mathcal{U}[0,1]$ & $0.25 _{ - 0.04 } ^ { + 0.05 } $ \\
    \hline
   \noalign{\smallskip}
  \end{tabular}
  \begin{tablenotes}\footnotesize
  \item \emph{Note} -- $^{(\mathrm{a})}$ $\mathcal{U}[a,b]$ refers to uniform priors between $a$ and $b$ and $\mathcal{F}[a]$ to a fixed value $a$. 
\end{tablenotes}
\end{table*}

\begin{figure*}
\centering
  \includegraphics[width=0.49\textwidth]{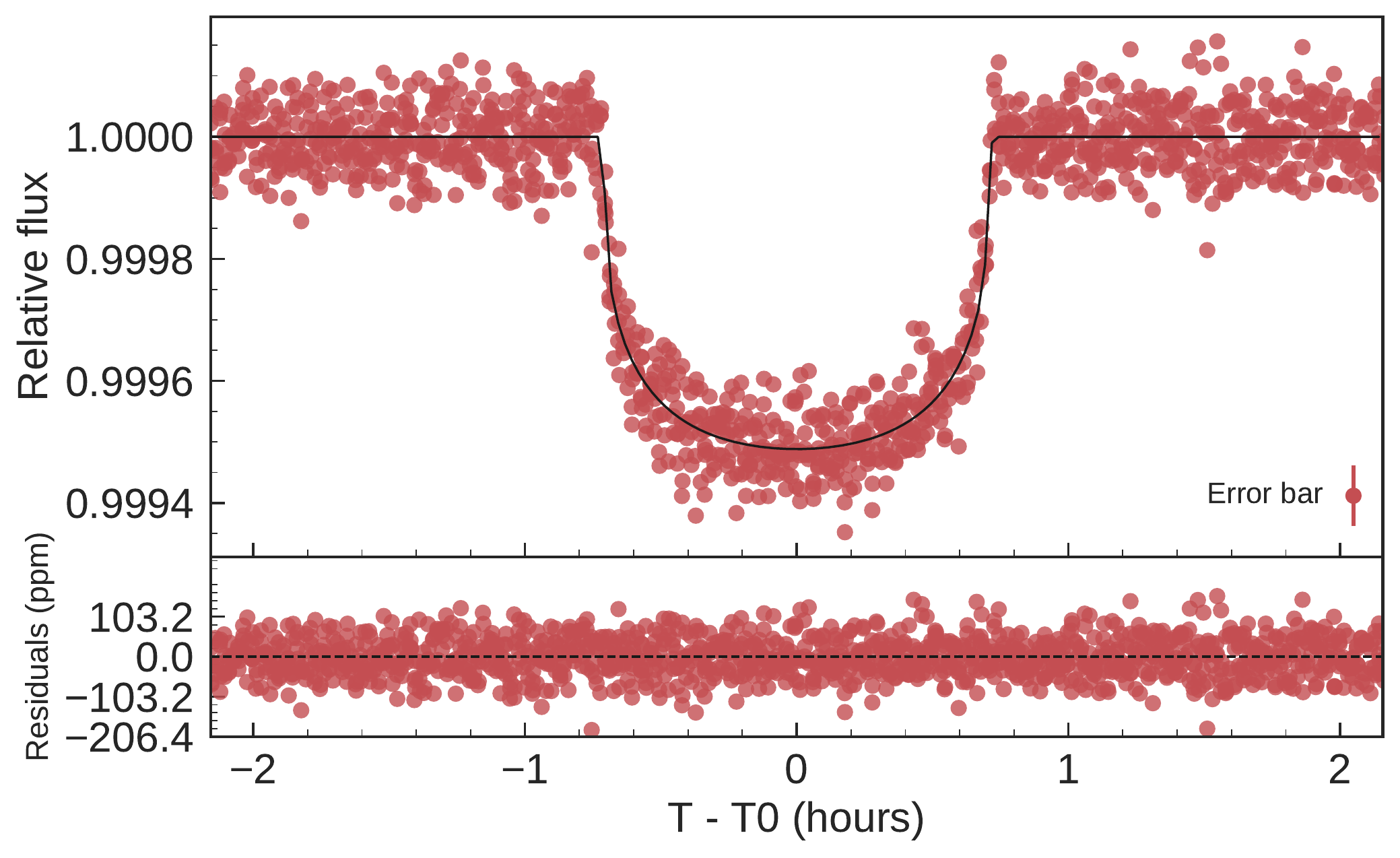}
  \includegraphics[width=0.49\textwidth]{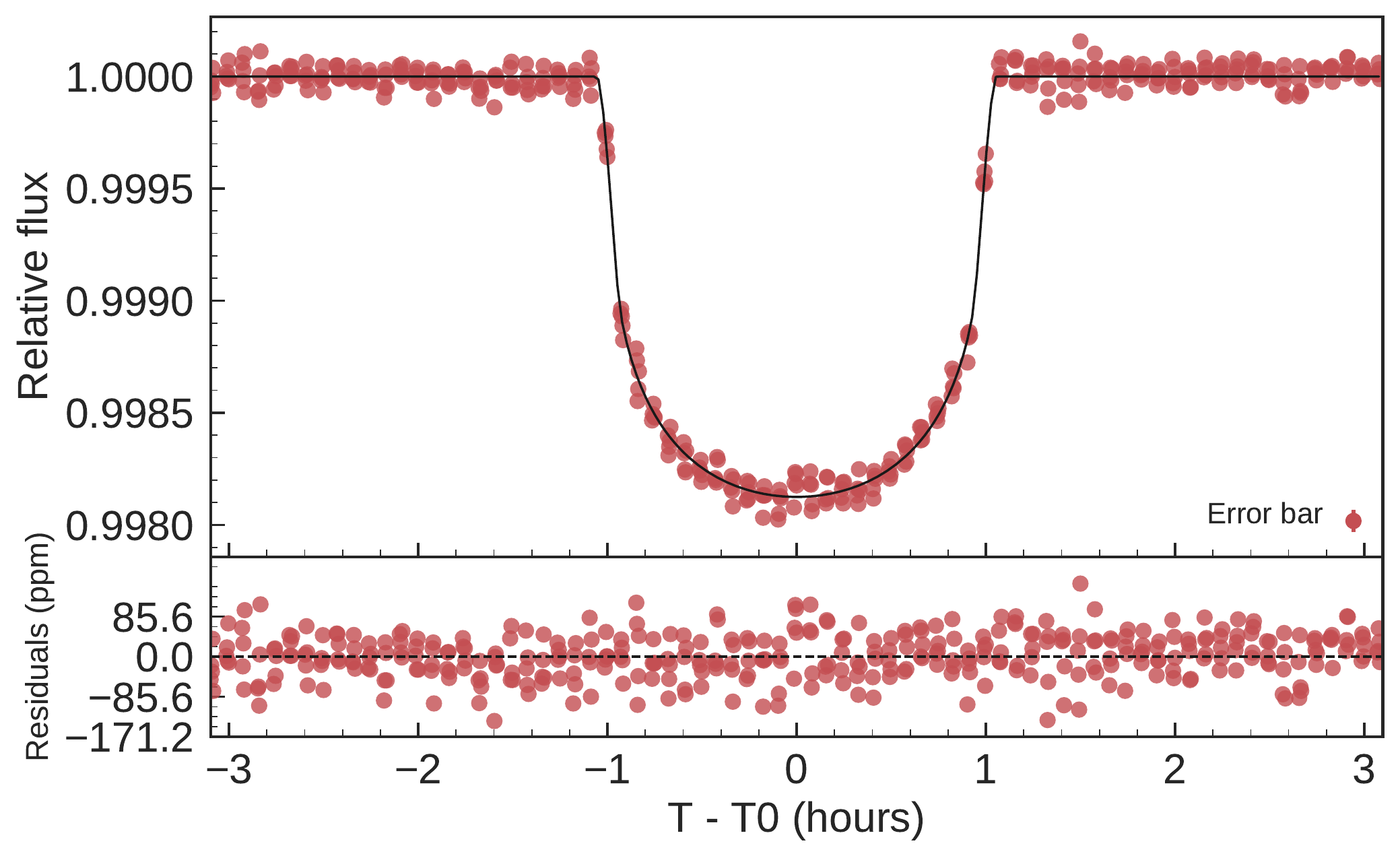}
  \includegraphics[width=0.49\textwidth]{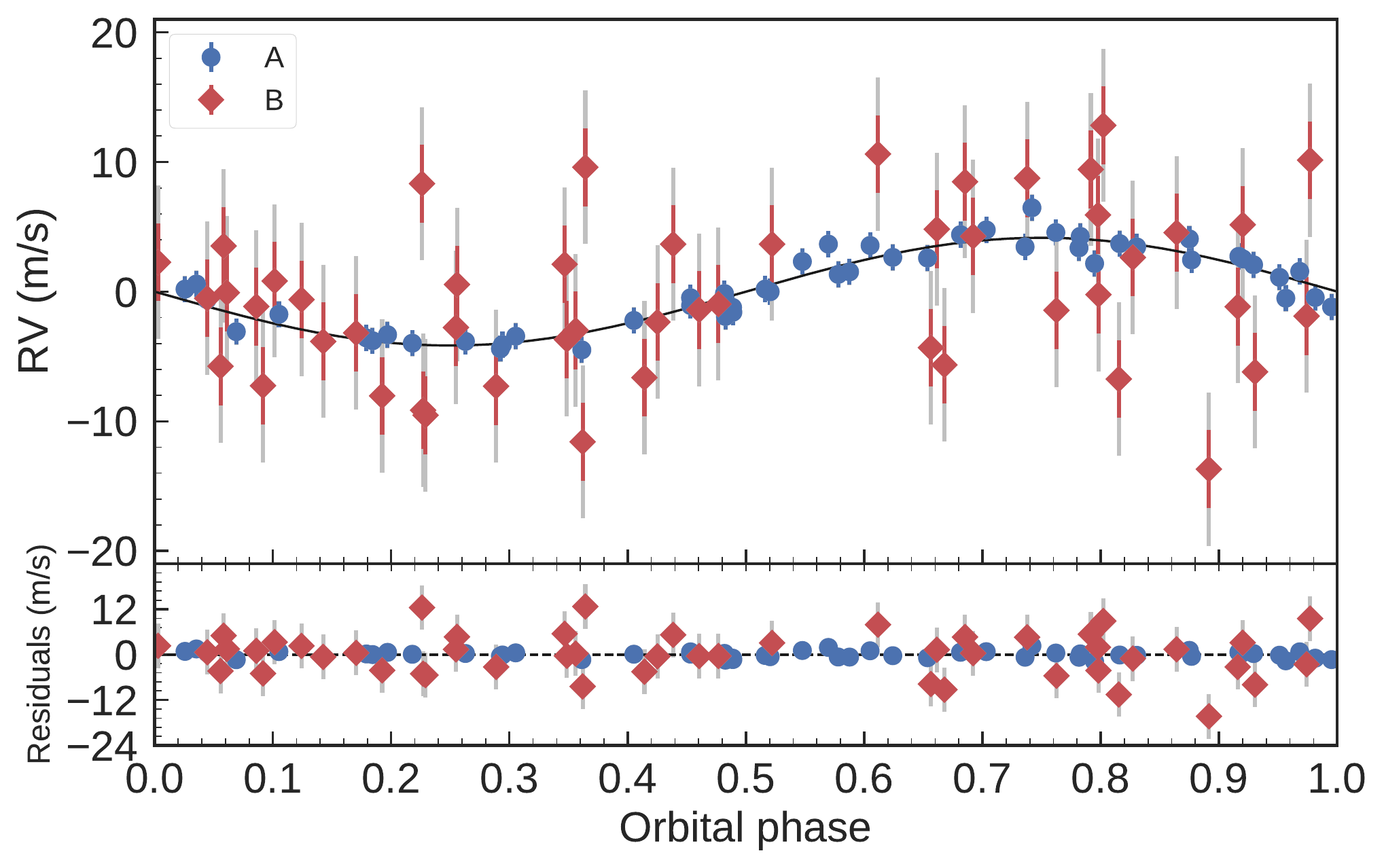}
  \includegraphics[width=0.49\textwidth]{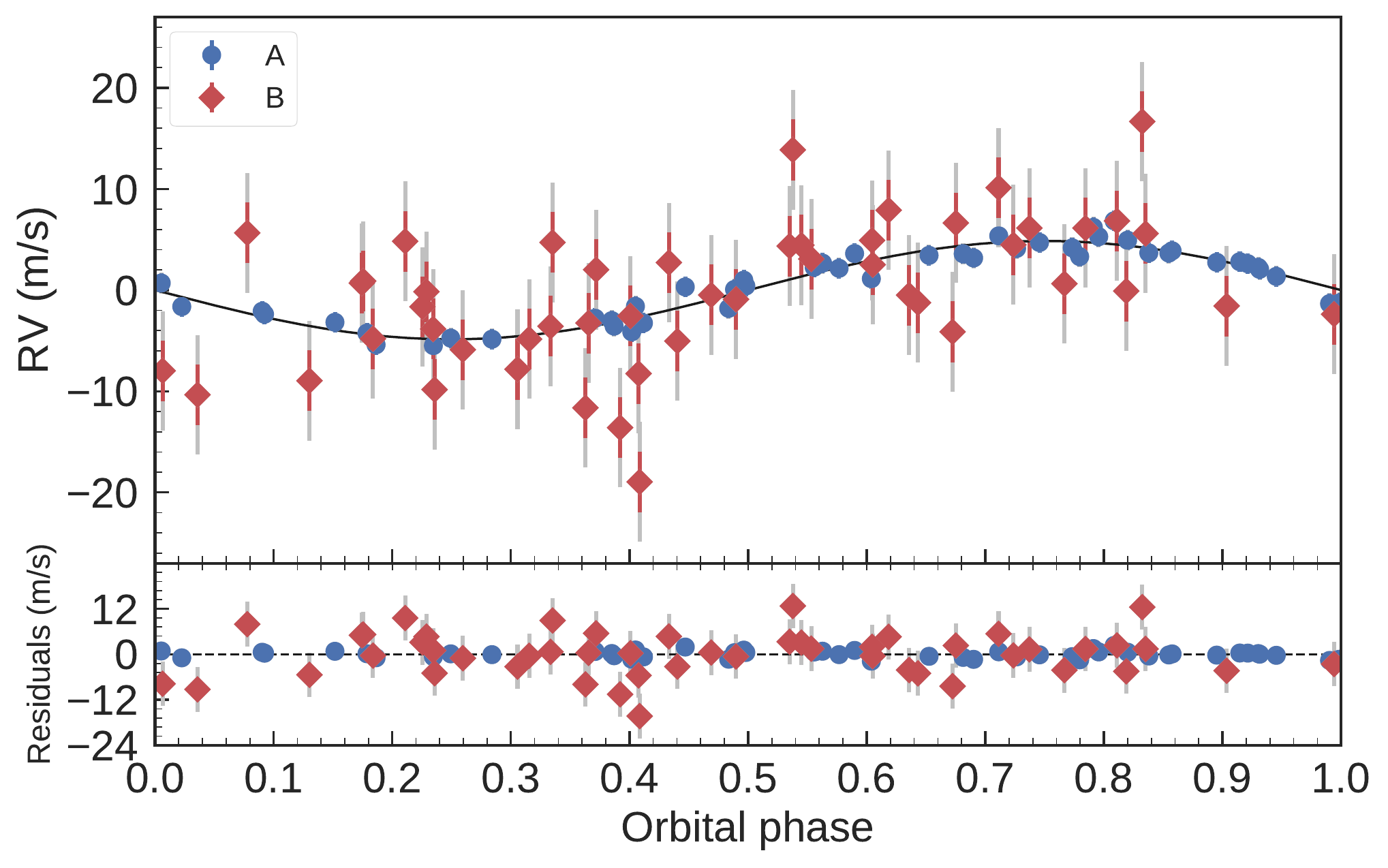} \\
  \includegraphics[width=0.49\textwidth]{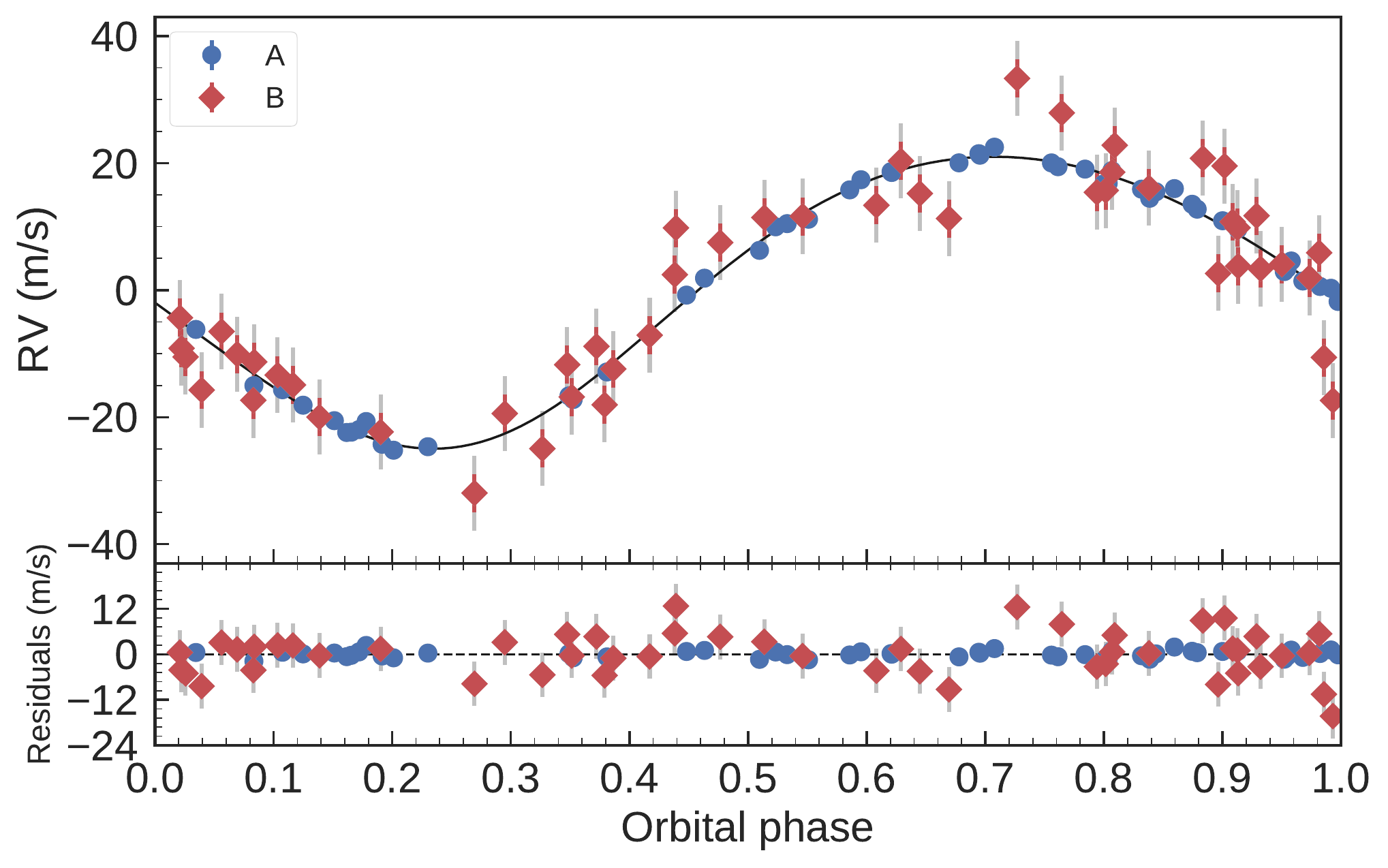}
  \caption{{\it Upper panels:} Phase-folded light curves for planet b and c of the simulated planetary system. Synthetic data points are plotted with the red circles. The best fitting transit models are over-plotted with thick black lines. {\it Middle and lower panels:} Phase-folded RV curves for planet b, c and d of the simulated planetary model. Synthetic data for instrument A is shown with blue circles, whereas for instrument B with red diamonds. The best fitting RV models are over-plotted with thick black lines. The gray error bars account for the jitter term for each instrument.
   \label{fig:modelfit1}}
\end{figure*}

\begin{figure*}
\includegraphics[width=0.95\textwidth]{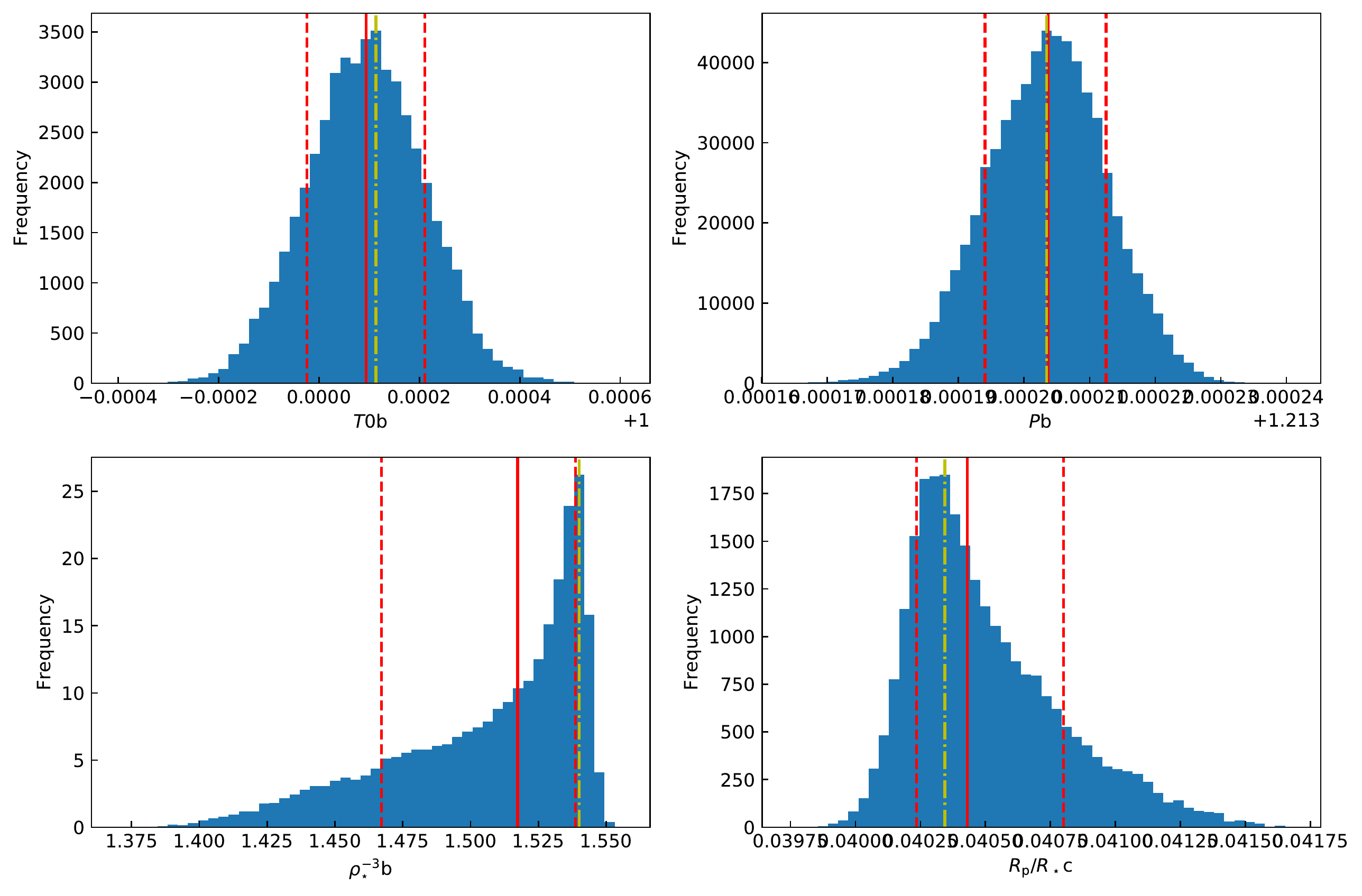}
\caption{Posterior distributions for $T_{0,{\rm b}}$ (upper left), $P_{\rm b}$ (upper right),  $\rho_{\star}^{1/3}$ (lower left), and $R_{\rm p, c}/R_\star$ (lower right) of the toy model fit. The solid red lines mark the medians of the distributions, whereas the dashed red lines mark the limits of the 68\% credible intervals. The mode are shown with the dot-dashed yellow lines.
\label{fig:posteriortest}}
\end{figure*}

The lower panel of Fig.\,\ref{fig:syndata} shows the simulated RV data and the inferred best-fitting three-planet model. Figure. \ref{fig:modelfit1} displays the phase-folded transit and RV curves. Figure\,\ref{fig:posteriortest} displays the posterior distributions of some of the fitted parameters. These histograms are useful diagnostic plots to check the goodness of the MCMC output. We note that our analysis provides unimodal posterior distributions. Their shapes are either Gaussian ($T_{0,{\rm b}}$ and $P_{\rm b}$), or skewed ($\rho_{\star}^{1/3}$ and $R_{\rm p,c}/R_{\star}$). The 68\% credible intervals are over plotted on each histogram; they corresponds to the error bars reported in Table\,\ref{tab:parameters}.

We note that the jitter term of instrument B found by \texttt{pyaneti} ($\sigma_{\rm B}\,=\,4.09_{-0.68}^{+0.77}$~\ms; cfr. Table\,\ref{tab:parameters}) agrees with the simulated value of 4\,\ms. We stress that jitter terms must be used when we have a reason to believe that the error bars are underestimated ($\chi^2/\mathrm{dof} > 1$). If a fit is poor, it is recommended to first check if the model can be improved before adding a jitter term. The derived stellar density is fully consistent with the density expected for the simulated host star. We emphasize that \texttt{pyaneti} fits for the stellar mean density instead for the scaled semi-major axis of planet b and c (see Sect.\,\ref{sec:parametrization}).

\subsection{The multi-planet system K2-38}
\label{sec:k238}

\subsubsection{Setup}

We also tested \texttt{pyaneti} with a real planetary system. We modeled the transit photometry and radial velocities of K2-38 and compared our results with those published by \citet{Sinukoff2016}. K2-38 is G2\,V star transited by two planets whose masses have been measured via Doppler spectroscopy. The inner planet, K2-38\,b, orbits the star every 4\,days. It has a mass of 12 $M_\oplus$ and a radius of 1.55 $R_\oplus$. The outer transiting planet, K2-38\,b, has an orbital period of 10.5 days, a mass of $9.8\,M_{\oplus}$, and a radius of $2.4\,R_\oplus$.

K2-38 was photometrically observed by the \texttt{K2} mission \citep{Howell2014} during its campaign 2. The RV measurements were gathered with the \texttt{HIRES} spectrograph \citep{Vogt1994} mounted at the Keck I 10 m telescope, at Keck Observatory (Mauna Kea, Hawai'i). \citet{Sinukoff2016} detected a linear trend in the RV measurements, indicative of the presence of an additional companion in the system. While modeling the RV data, the authors added a jitter term to the nominal uncertainties to account for instrumental velocity noise not included in the nominal uncertainties and/or possible sources of stellar variability. Because of its complexity, this system is an ideal test-bench for \texttt{pyaneti}.

We used the EVEREST processed light curve \citep{Luger2016} to perform the transit light curve analysis. We de-trended the \texttt{K2} data with \texttt{exotrending} \citep{exotrending} by fitting a second-order polynomial function to the 5-hour out-out-transit data centered around each transit. The RV measurements were taken from \citet{Sinukoff2016}.

We used the general form of the likelihood given in eq.~(\ref{eq:likelihooddatum}) to account for the RV jitter term. We added a linear trend term $\dot{\gamma}$ to equation (\ref{eq:rvmulti}) taking as zero point the time of conjunction of planet b. We super-sampled the transit model by a factor of 10 to account for the \emph{K2} long-cadence data \citep{Kipping2013}. We fixed $q_2$ to 0.5 to recover the linear limb darkening case and set {\bf Gaussian} priors on $q_1$ with 1-$\sigma$ uncertainty of 0.1. We set uniform priors for the remaining parameters (details are provided in Table\,\ref{tab:parsk238}) and assumed circular orbits as adopted by \citet{Sinukoff2016}. The sampling of the parameter space follows the procedure described in Sect.\,\ref{Synthetic_data}. Briefly, we initialized 100 independent chains created randomly inside the prior ranges. Once all chains converged, we created posterior distributions with 50,000 independent points for each parameter.

\subsubsection{Results}

\begin{figure*}
  \includegraphics[width=\textwidth]{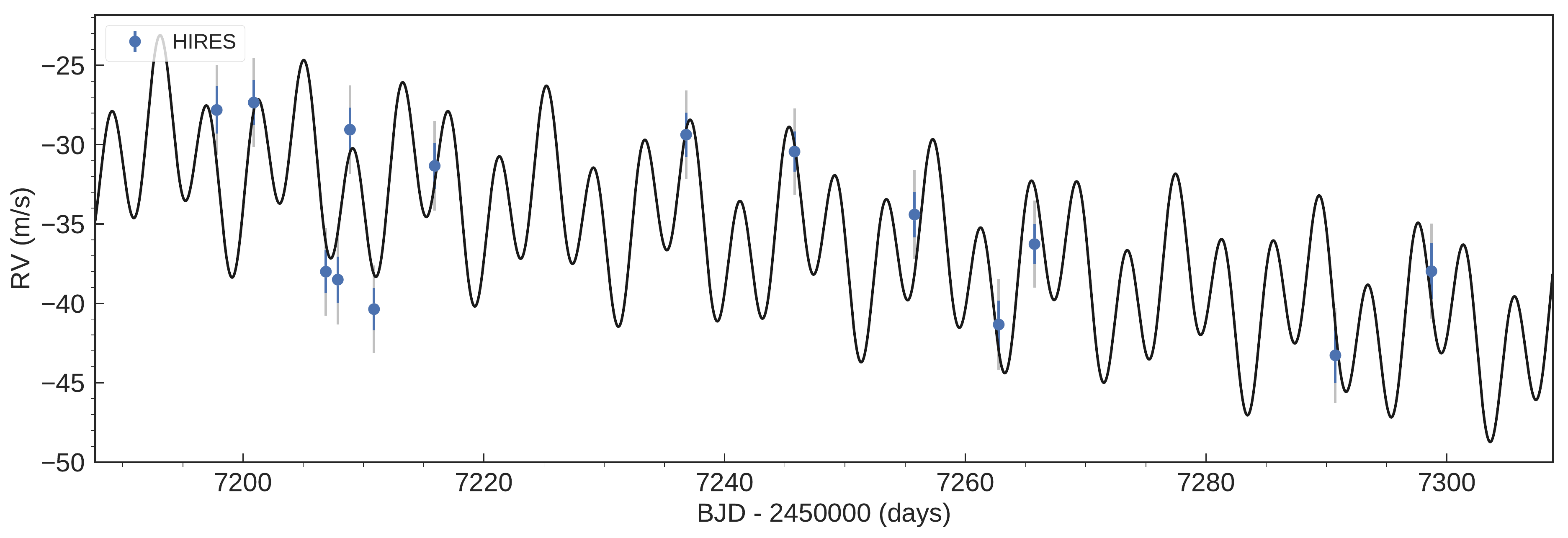}\\
  \includegraphics[width=0.49\textwidth]{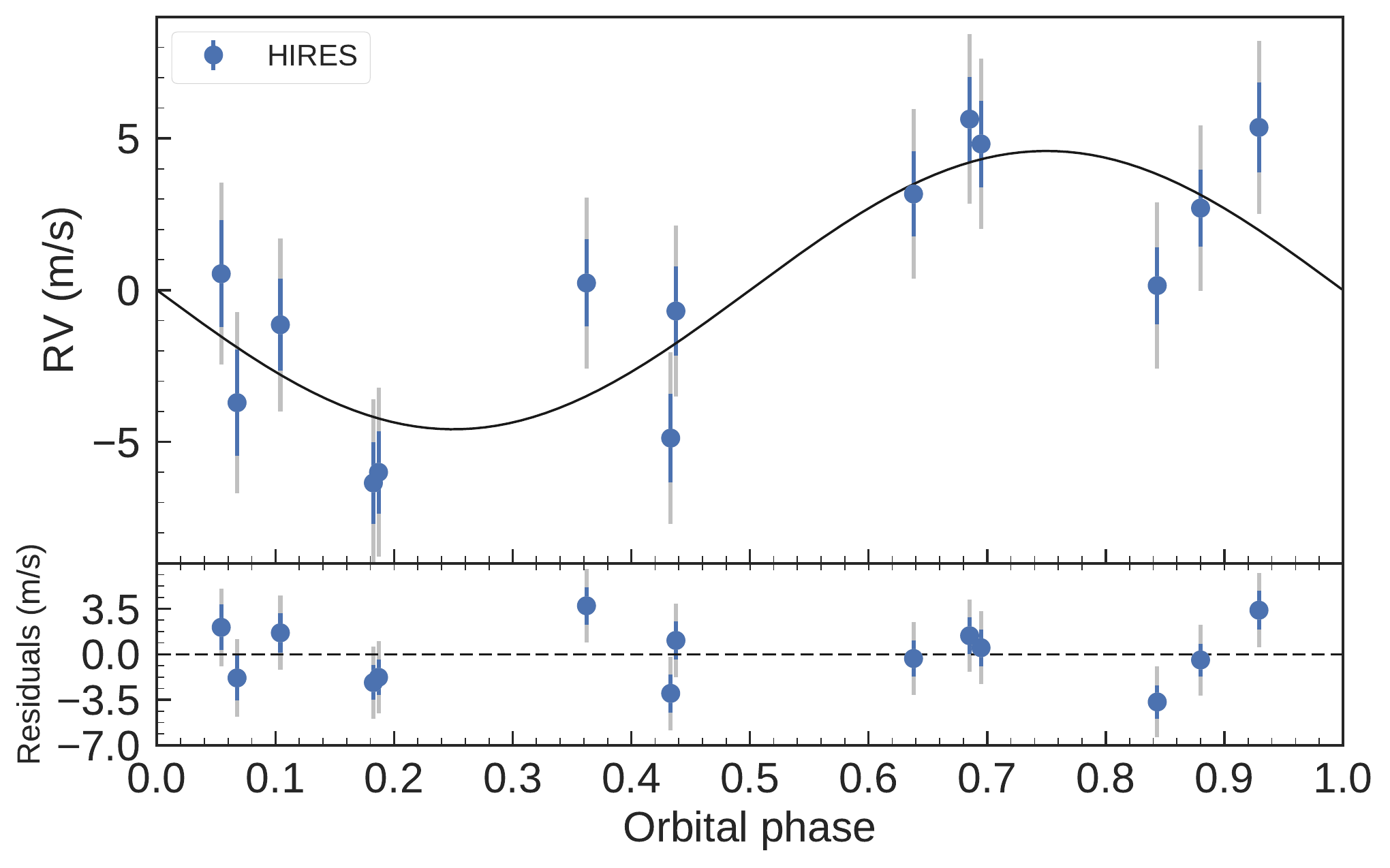}
  \includegraphics[width=0.49\textwidth]{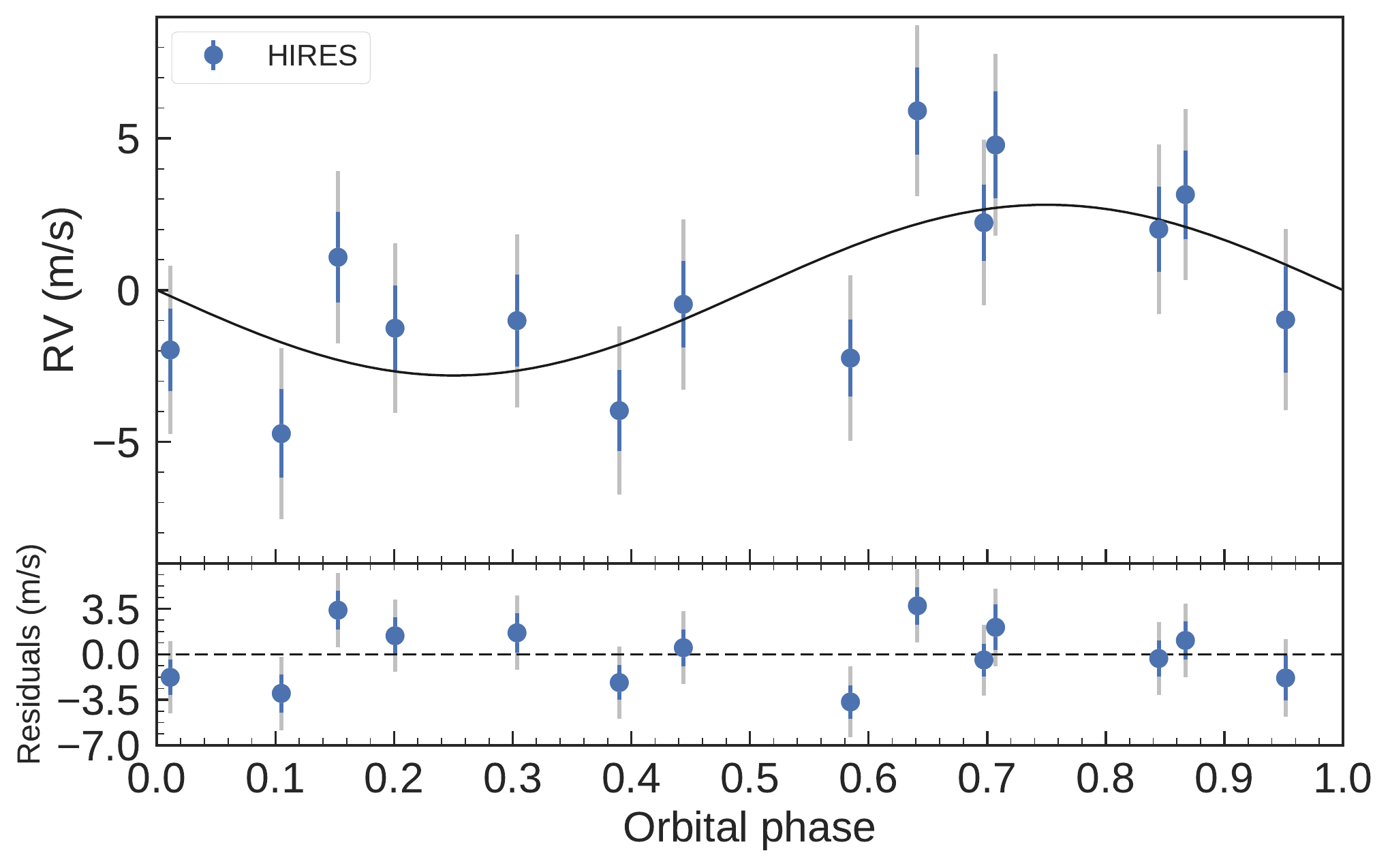}\\
  \includegraphics[width=0.49\textwidth]{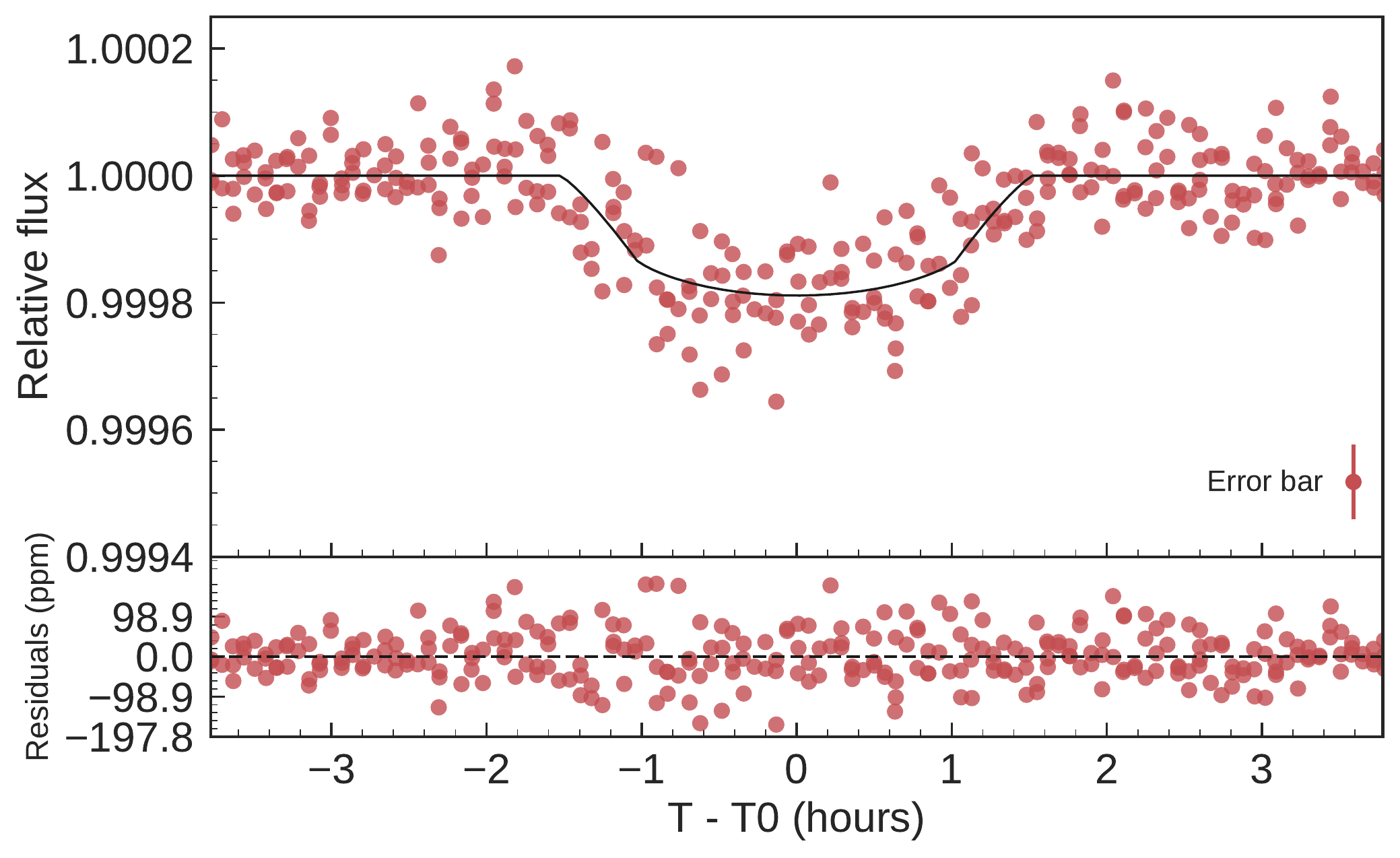}
  \includegraphics[width=0.49\textwidth]{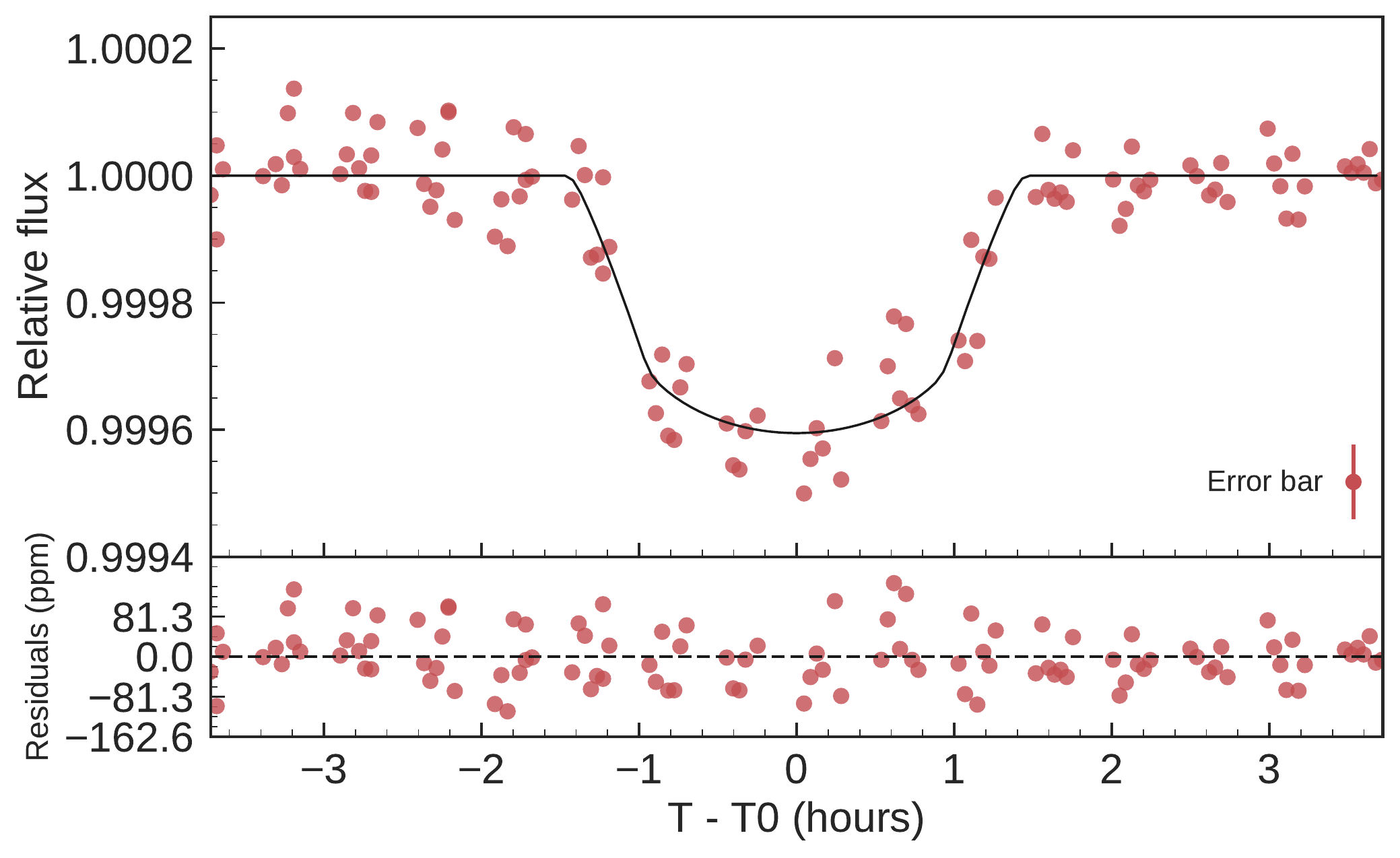}\\
  \caption{ \textit{Upper panel}: HIRES RV measurements of K2-38 (blue circles). The best-fitting solution is shown with a thick black line. A linear trend is visible in the data. The fitted stellar jitter is shown as a gray extension to the nominal error bars. \textit{Middle panels}: Phase-folded RV curves for K2-38\,b (left) and K2-38\,c (right). \textit{Lower panels}: Transit light curve folded to the orbital period of K2-38\,b  (left) and K2-38\,c (right), and residuals. The red points are \texttt{K2} data. The thick black lines mark the best fitting transit models. \label{fig:fitsk238}}
\end{figure*}

The final estimates and their 1-$\sigma$ uncertainties are taken as the median and the 68\,\% of the credible interval of the posterior distributions. Values are reported in Table\,\ref{tab:parsk238}. Photometric and RV data, along with the best fitting transit and RV models are displayed in Figure \ref{fig:fitsk238}.

We compare our results with those from \citet{Sinukoff2016} in Table \ref{tab:parsk238}. The parameter estimates agree well within their 1-$\sigma$ uncertainties. However, we note that the largest discrepancies are found for the parameters derived from the \texttt{K2} data. This is very likely due to the different extracted light curve used in our analysis, as well as on the different transit de-trending algorithm. As for the RV-derived parameters, our results are in excellent agreement with those reported by \citet{Sinukoff2016}.

\begin{table*}
  \caption{System parameters. \label{tab:parsk238}}  
  \begin{tabular}{lccc}
  \hline
  Parameter & \citet{Sinukoff2016} & Prior$^{(\mathrm{a})}$ & Fitted value \\
  \hline
  \multicolumn{4}{l}{\emph{Model Parameters planet b}} \\
  \noalign{\smallskip}
    Orbital period $P_{\mathrm{orb}}$ (days) & $4.01593 \pm 0.00050$ &  $\mathcal{U}[ 4.0134 , 4.0184 ]$ & $4.01632 _{ - 0.00034 } ^ { + 0.00032 } $  \\
    Transit epoch $T_0$ (BJD - 2,450,000) & $6896.8786 \pm 0.0054$ & $\mathcal{U}[  6896.8486 , 6896.9086  ]$ & $6896.8734 _{ - 0.0034 } ^ { + 0.0038 } $  \\  
    Scaled semi-major axis $a/R_{\star}$ & $10.7_{-3.7}^{+1.3}$ &  $\mathcal{U}[ 1.1 , 50 ]$ & $11.3_{ - 2.3 } ^ { + 1.0 } $\\
    Scaled planet radius $R_\mathrm{p}/R_{\star}$ & $0.01281^{+0.00105}_{-0.00064}$ & $\mathcal{U}[0,0.1]$ & $0.01247 _{ - 0.00045 } ^ { + 0.00087 } $  \\
    Impact parameter, $b$ & 0.$48 \pm 0.30$ & $\mathcal{U}[0,1]$  & $0.35_{ - 0.25 } ^ { + 0.33 } $ \\
    $e $ & 0 &  $\mathcal{F}[0]$ & 0  \\
    $\omega_\star$ (deg)  & 90 & $\mathcal{
    F}[90]$ & 90 \\
    Radial velocity semi-amplitude variation $K$ (m s$^{-1}$) & $4.6 \pm 1.1$ & $\mathcal{U}[0,100]$ & $4.6 \pm 1.1$ \\
  \hline
  \multicolumn{4}{l}{\emph{Model Parameters planet c}} \\
  \noalign{\smallskip}
    Orbital period $P_{\mathrm{orb}}$ (days) & $10.56103 \pm 0.00090$ &  $\mathcal{U}[ 10.5565 , 10.5655]$ & $10.56155 \pm 0.00049 $ \\
    Transit epoch $T_0$ (BJD - 2,450,000) & $6900.4752 \pm 0.0033$ & $\mathcal{U}[ 6900.4552 , 6900.4952 ]$ & $6900.4740 \pm 0.0018 $  \\  
    Scaled semi-major axis $a/R_{\star}$ & $26.3_{-16.1}^{+5.4}$ &  $\mathcal{U}[1.1,50]$ &$31.3 _{ - 5.1 } ^ { + 2.1 } $ \\
    Scaled planet radius $R_\mathrm{p}/R_{\star}$ & $0.02004^{+0.0024}_{-0.0013}$   & $\mathcal{U}[0,1]$ & $0.01841 _{ - 0.0005 } ^ { + 0.0010 } $  \\
    Impact parameter, $b$ & $0.64^{0.23}_{-0.41}$ & $\mathcal{U}[0,1]$  & $0.34 _{ - 0.25 } ^ { + 0.27} $ \\
   $e $ & 0 &  $\mathcal{F}[0]$ & 0  \\
    $\omega_\star$ (deg)  & 90 & $\mathcal{
    F}[90]$ & 90 \\
    Radial velocity semi-amplitude variation $K$ (m s$^{-1}$) & $2.8 \pm 1.3$ & $\mathcal{U}[0,1000]$ & $2.8 \pm 1.3 $ \\
      \hline
      \multicolumn{4}{l}{\emph{Other Parameters}} \\
  \noalign{\smallskip}
      RV value at $T_{0,1}$ $\gamma$ (\ms) $^{(\mathrm{b})}$ & $-1.7 \pm 0.9 $  & $\mathcal{U}[-1000,1000]$ & $0.034 \pm 0.010$ \\
      Linear trend slope $\dot{\gamma}$ (\kms\,d$^{-1}$) & $-0.101 \pm 0.030$  & $\mathcal{U}[-1,1]$ & $-0.103 \pm 0.029$ \\
      HIRES jitter term $\sigma_{\rm HIRES}$ (\ms) & $2.4^{+1.0}_{-0.7}$  & $\mathcal{U}[0,1000]$ & $2.4 ^{+ 1.0}_{- 0.7}$ \\
    Parameterized limb-darkening coefficient $q_1$ & $0.38 \pm 0.1$ $^{(\mathrm{c})}$ & $\mathcal{N}[0.38,0.1]$ & $0.42 \pm 0.1 $ \\
    Parameterized limb-darkening coefficient $q_2$ & 0.5 $^{(\mathrm{c})}$ & $\mathcal{F}[0.5]$ & 0.5 \\
    \hline
   \noalign{\smallskip}
  \end{tabular}
  \begin{tablenotes}\footnotesize
  \item \emph{Note} -- $^{(\mathrm{a})}$ $\mathcal{U}[a,b]$ refers to uniform priors between $a$ and $b$, $\mathcal{N}[a,b]$ to Gaussian priors with median $a$ and standard deviation $b$, and $\mathcal{F}[a]$ to a fixed value $a$. 
$^{(\mathrm{b})}$ Our results and \citeauthor{Sinukoff2016} results do not agree because the instant at which the intercept is calculated is not the same. $^{(\mathrm{c})}$ We transform the values reported by \citeauthor{Sinukoff2016} to the $q_1$ and $q_2$ parametrization to perform the comparison.  
\end{tablenotes}
\end{table*}

This test confirms the correct {\bf implementation} of the MCMC method and multiplanet equations. \citet{Sinukoff2016} used the widely used ensemble sampler package \texttt{emcee} \citep{Foreman2013}. 

\subsection{Execution performance}

We show here that \texttt{pyaneti} is able to produce scientific results within a few minutes in a personal laptop. We ran the test case presented in section \ref{sec:k238} (2 planet system, 435 data points, 10\,000 iterations with 100 independent Markov chains) with different CPU configurations. We used a machine with an Intel i7-6500U CPU (Four 2.50GHz cores) and with Linux (Fedora 64-bit) operating system. We compiled the code with \texttt{gfortran}\,8.1.1 and used 1, 2, and 4 CPUs. The respective execution times were 10m\,11s, 5m\,56s and 4m\,22s. These results prove the power of the code to perform a full run in a personal laptop. However, we stress that the execution time depends on the analyzed data set. Based on our experience with \texttt{pyaneti}, the modeling of only RV data is carried withing a few minutes. For demanding fits requiring longer execution time (e.g., long time-series photometry), \texttt{pyaneti} can be ran in parallel in a server machine equipped with more than one CPU.

\section{Conclusions}
\label{sec:conclusions}

We have developed and tested the code \texttt{pyaneti}, a software suite able to simultaneously fit RV and transit light curves of multi-planet systems. \texttt{pyaneti} combines the computational power of \fortran\ with the versatility of \python\ and it offers the option to run in parallel with \texttt{OpenMP}. The package has been developed under ``the open source ideology'', i.e., both the code and the platforms used to write the package are totally free.

We have tested \texttt{pyaneti} with synthetic data and proved that the code is able to recover the parameters of multi-planet systems. We have also performed an independent fit of K2-38 and our results are consistent with those in the literature. The joint modeling of the transit and RV measurements of the K2-38 system takes only $\sim$5\,min on a personal laptop. This demonstrates that the code can perform fast analyses and makes \texttt{pyaneti} a powerful tool to perform data analysis of hundreds of systems coming from future space- and ground-based instruments, such as \texttt{TESS}, \texttt{PLATO}, \texttt{CHEOPS}, and \texttt{ESPRESSO}.

Future releases of \texttt{pyaneti} will include extra parametric models, such as TTV, multi-band photometry and phase curves. We anticipate that the code will also be able to use other likelihoods and priors, such as Gaussian processes.

\section*{Acknowledgements}

D.~Gandolfi gratefully acknowledges the financial support of the \emph{Programma Giovani Ricercatori -- Rita Levi Montalcini -- Rientro dei Cervelli (2012)} awarded by the Italian Ministry of Education, Universities and Research (MIUR).



\bibliographystyle{mnras}
\bibliography{bibs} 




\appendix

\section{Numerical treatment of the posterior}
\label{ap:ap1}

Equation (\ref{eq:likelihood}) may lead to very small/big numbers which generate numerical overflows. Therefore it is convenient to use the logarithmic of probability densities. Bayes' theorem is rewritten as

\begin{equation}
\ln P(M|D) = \ln P(D|M) + P(M) - P(D).
\label{eq:loglikelihood}
\end{equation}

We note that this treatment of the posterior does not affect the MCMC method. Since the ratio between the actual and proposed states can be calculated easily as

\begin{equation}
\frac{P(D|\vec{\Phi})P(\vec{\Phi})}{P(D|\vec{\phi})P(\vec{\phi})} =
\exp
\left[
\ln P(D|\vec{\Phi}) + \ln P(\vec{\Phi}) -
\ln P(D|\vec{\phi}) - \ln P(\vec{\phi})
\right]
\end{equation}

By following this approach, the general form of the Gaussian likelihood for an RV and transit fit is given using eq. (\ref{eq:likelihood}) as

\begin{equation}
P(D|M) = 
\prod_i^{N_{\rm RV}} \left[ \frac{1}{\sqrt[]{ 2\pi (\sigma^2_i + \sigma^2_{\rm j})} } \right]_{\rm RV}
\times
\prod_i^{N_{\rm LC}} \left[ \frac{1}{\sqrt[]{ 2\pi (\sigma^2_i + \sigma^2_{\rm j})} } \right]_{\rm LC} 
 \times
\exp \left\{ - \frac{1}{2} \chi^2_{\rm Tot} \right\}
,
\end{equation}

where

\begin{equation}
\chi^2_{\rm Tot} = 
\sum_i^{N_{\rm RV}}
\frac{ (D_{i,{\rm RV}} - M_{i,{\rm RV}})^2 } {\sigma_i^2 + \sigma_{\rm j}^2} +
\sum_i^{N_{\rm LC}}
\frac{ (D_{i,{\rm LC}} - M_{i,{\rm LC}})^2 } {\sigma_i^2 + \sigma_{\rm j}^2}.
\end{equation}

The ${\rm RV}$ and ${\rm LC}$ sub-indexes refers to RV and light curve data and models, respectively. The logarithmic form of the likelihood given in eq. (\ref{eq:likelihood}) is rewritten as

\begin{equation}
\ln P(D|M) = 
-\frac{1}{2}
 \left[
 \sum_i^{N_{\rm RV}} \ln  2\pi (\sigma^2_i + \sigma^2_{\rm j}) 
 +
  \sum_i^{N_{\rm LC}} \ln  2\pi (\sigma^2_i + \sigma^2_{\rm j}) 
 +
 \chi^2_{\rm Tot}
 \right].
 \label{eq:likelihoodtot}
\end{equation}

We note that eq. (\ref{eq:likelihoodtot}) can be used too model pure RV or transit data.




\bsp	
\label{lastpage}
\end{document}